\documentclass[prd,a4paper,twocolumn,nofootinbib]{revtex4}
 \usepackage{mathrsfs}
\usepackage{graphicx}
\usepackage{latexsym}
\usepackage{amsmath}
\usepackage{amssymb}
\usepackage{latexsym}
\usepackage{textcomp}
\usepackage{amsbsy}
\usepackage{graphics}
\usepackage{braket}
\usepackage{xcolor}
\usepackage{calc}
\usepackage{enumitem}
\newlength{\depthofsumsign}
\begin{document}
\title{Lattice Gauge Theories and  Spin Models}
\author{Manu Mathur}
\email{manu@bose.res.in}
\affiliation{S. N. Bose National Centre for Basic Sciences, Salt Lake, JD Block, Sector 3, Kolkata 700098, India} 
\author{T. P. Sreeraj}  
\email{sreerajtp@bose.res.in}
\affiliation{S. N. Bose National Centre for Basic Sciences, Salt Lake, JD Block, Sector 3, Kolkata 700098, India} 
\begin{abstract}
\noindent 
The Wegner $Z_2$ gauge theory-$Z_2$ Ising spin model duality in $(2+1)$ dimensions  is revisited and derived through a series of canonical transformations. The Kramers-Wannier duality is similarly obtained.
The Wegner $Z_2$ gauge-spin duality is directly generalized to SU(N) lattice gauge theory in $(2+1)$ dimensions to obtain the SU(N) spin model in terms of  the SU(N) magnetic fields and their conjugate
SU(N) electric scalar potentials.  
The exact \& complete solutions of the 
$Z_2, U(1), SU(N)$ Gauss law constraints in terms of the 
corresponding spin or dual potential operators are given. 
The gauge-spin duality  naturally leads to a new 
gauge invariant magnetic disorder operator for SU(N)
lattice  gauge theory which produces a magnetic vortex 
on the plaquette.   
A variational ground state of the  
SU(2) spin model with nearest neighbor interactions is constructed to analyze SU(2) gauge theory. 
\end{abstract}
\maketitle
\section{Introduction}
In 1971 Franz Wegner, using duality transformations, showed  that   in two space dimensions
$Z_2$ lattice gauge theory  can be exactly mapped into a $Z_2$ Ising model describing  spin half  magnets \cite{wegner}. This is the earliest and the simplest example of the intriguing gauge-spin duality.  
Wegner's work, in turn,  was strongly motivated by the self-duality of planar Ising model discovered by Kramers and Wannier 30 years earlier \cite{km}. 
Such alternative dual descriptions have been 
extensively discussed in the past  as they  are  useful to understand theories and their phases at a deeper level  \cite{thmds,hooft,banks,fradsuss,kogrev,savit,horn,dualsup,baal,sharatram,manu}. 
In the context of QCD, the duality transformations 
have been studied  to understand color confinement via dual superconductivity \cite{thmds,hooft,dualsup,baal} and to extract topological degrees of freedom \cite{banks,savit,sharatram,manu}. 
They may relate the important and relevant 
degrees of freedoms at high and low energies 
providing a better understanding of  
non-perturbative issues  in low energy  QCD.  
The duality ideas in the Hamiltonian framework 
are also relevant for the  recent quest to build quantum simulators  for abelian and non-abelian 
lattice gauge theories using cold atoms in optical lattices \cite{reznik}. 
In these cold atom experiments, the (dual) spin description of SU(N) lattice  gauge theory should be useful as there are no exotic, quasi-local Gauss law constraints to be implemented at every lattice site \cite{coldatomgl}. The duality methods and the resulting spin models, without redundant local gauge degrees of freedom, can also provide  more efficient tensor network, variational ansatzes for the low energy states  of SU(N) lattice gauge theories \cite{tn}.   

In this work, we start with a  brief overview of  Kramers-Wannier and 
Wegner dualities within the Hamiltonian framework. We show that these old, well-established spin-spin and  gauge-spin dualities 
can be  constructively obtained through a series of iterative canonical transformations.
These canonical transformation techniques are easily generalized to SU(N) lattice gauge theory to obtain the equivalent dual  SU(N) spin model without any local gauge degrees of freedom. Thus using canonical transformations we are able to treat spin, abelian and non-abelian dualities 
on the same footing. 
In $(2+1)$ dimensions the  spin operators in the dual spin models  
are the  scalar $Z_2, U(1), SU(N)$ magnetic fields and 
their conjugate electric scalar potentials respectively. These spin operators  solve the $Z_2, 
U(1)$ and $SU(N)$ Gauss laws. 

The Kramers-Wannier and Wegner dualities naturally 
lead to construction of disorder operators and order-disorder algebras 
\cite{kogrev,fradsuss,horn}. In both cases the disorder operators are simply the dual spin operators creating  $Z_2$ kinks and $Z_2$ magnetic vortices on  plaquettes respectively. Note that these
creation operators are highly non-local in terms of the operators of the  
original Ising model or $Z_2$ gauge theory. 
Therefore without duality transformations 
they are  difficult to guess.  We generalize these elementary 
duality ideas  to non-abelian gauge theories
after  briefly recapitulating them in the   
simpler $Z_2$ contexts mentioned above. In particular, we exploit SU(N) 
dual spin operators  to construct a new gauge 
invariant disorder operator for SU(N) 
lattice gauge theory. 
Further, like in $Z_2$ lattice gauge theory, 
the  non-abelian order-disorder algebra 
involving SU(N) Wilson loops and SU(N) 
disorder operators is worked out.
The interesting role of non-localities 
in non-abelian duality  resulting in the solutions of SU(N) Gauss laws and production of local vortices is discussed.  
For the sake of clarity and continuity, the SU(N) gauge theory results will always be discussed in the background of the corresponding  Ising model, $Z_2$ gauge theory results. The similar features amongst them are  emphasized and the differences are also pointed out.   


In the context of
$Z_2$ gauge theory in $(2+1)$
 dimensions, the two 
essential features of Wegner  
duality  \cite{wegner} are 
\begin{itemize} 
\item it eliminates all unphysical gauge
degrees of freedom mapping it into $Z_2$ spin  model  with a  $Z_2$ global symmetry. {\it There are no $Z_2$ Gauss law constraints in the dual $Z_2$ spin model}.

\item it  maps the interacting (non-interacting) terms in the $Z_2$  lattice gauge theory Hamiltonian into  non-interacting (interacting) terms in the $Z_2$ 
spin  model Hamiltonian resulting in the inversion of the coupling constant. 

\end{itemize}

It is important to note that the above gauge-spin duality 
is through the  loop description of $Z_2$ lattice 
gauge theory. The original Hamiltonian is written 
in terms of fundamental $Z_2$ electric fields and their conjugate magnetic vector potentials. The $Z_2$ magnetic fields are not fundamental and  
obtained from magnetic 
vector  potentials.    
On the other hand, in the dual Ising model the fundamental 
spin degrees of freedom  are the  
$Z_2$ magnetic 
fields and their conjugate  electric scalar potentials.  
Now the  electric fields are not fundamental and are obtained from the 
electric scalar potentials.
 We arrive at this  dual spin description through a series of canonical transformations. They   
 convert the initial electric fields, magnetic 
vector potentials  
into the following two mutually independent {\it physical} \&  {\it unphysical } classes of operators: 
\begin{enumerate} 
\item {\it  $Z_2$ spin or plaquette loop operators}: 
representing the {\it physical} $Z_2$ magnetic fields and their conjugate electric scalar potentials over the plaquettes (see Figure \ref{dual}-a), 
\item {\it  $Z_2$ string  operators}: representing the $Z_2$ electric fields and the $Z_2$ flux operators of the {\it unphysical} string degrees of freedom. These strings isolate all  $Z_2$ gauge  degrees of freedom (see Figure \ref{dual}-b). 
 \end{enumerate} 
The interactions of spins in the first set are 
described by Ising model.
The corresponding physical Hilbert space is denoted by ${\cal H}^p$. The  second complimentary set, containing $Z_2$ string operators, represents all possible redundant gauge degrees of freedom. We show that 
the Gauss law constraints freeze all strings 
leading  to 
the Wegner gauge-spin duality within  ${\cal H}^p$.  
Further,  
the electric scalar  potentials are shown to be the solutions of the  $Z_2$ Gauss law constraints.
{\it Note that no gauge fixing is required  to obtain the dual description.}  
We show that the above  duality features also 
remain valid when these canonical/duality transformations are generalized to SU(N) 
lattice gauge theory.
As in the $Z_2$ case, the SU(N) Kogut-Susskind link operators get transformed  into the {\it physical 
spin/loop} and {\it unphysical  string} 
operators. Again the SU(N) strings are frozen and 
the  dual SU(N) spin operators provide all solutions 
of SU(N) Gauss law constraints.
In fact, these SU(N) canonical transformations have 
been discussed earlier in the context of 
loop formulation of SU(N) lattice gauge 
theories  \cite{msplb,msprd}. The motivation was to address the issue  of redundancies of Wilson loops or equivalently solve the SU(N) Mandelstam 
constraints. 
We now exploit them   
in the context of non-abelian duality. 

The plan of the paper is as follows.
In section \ref{sgsdct}, we discuss the  
canonical transformation techniques  
to systematically obtain Kramers-Wannier, Wegner and then SU(N)  dualities.  The compact U(1) lattice gauge theory duality can also be easily obtained from the SU(N) duality by 
ignoring the non-abelian, non-local terms. In section \ref{skwdt}, 
we start with the simplest Kramers-Wannier duality in the Ising model.
The  order-disorder 
operators, their algebras and 
creation, annihilation  of kinks 
are briefly 
discussed for the sake of uniformity and later 
comparisons \cite{km,kadanoff,fradsuss,kogrev,savit,horn}.  In section \ref{sz2gtsm}, we extend these 
canonical transformations to discuss Wegner
duality  in $2+1$ dimensions. We again obtain 
the old and well established results \cite{fradsuss,kogrev,savit,wegner,horn}  
with canonical transformations as the 
new ingredients. The $Z_2$ gauge theory 
order-disorder operators, their algebras and $Z_2$ 
magnetic vortices  are briefly summarized.  
In section \ref{snalgt}, the  $Z_2$ canonical transformations are generalized to SU(N) lattice  gauge theories leading to a SU(N) spin  model. As mentioned before  the SU(N) discussions 
are  parallel to the  $Z_2$ discussions for clarity.
A comparative summary of $Z_2$ gauge-spin and SU(N) gauge-spin operators is given in Table 1.  
At the end of section \ref{snalgt}, 
we construct the new SU(N) 
disorder operator.
The special case of 't Hooft disorder operator 
is discussed.  
The Wilson-'t Hooft loop algebra is derived.
The last section \ref{vgs} is devoted to variational analyses of the truncated SU(N) spin model. 
A simple `single spin' variational  ground state of the dual SU(N) spin model is constructed. The Wilson loop  in this ground state is shown to have area law behavior. 
In Appendix \ref{appa}, we discuss  explicit 
constructions of $Z_2$ Wegner duality through 
canonical transformations. 
In Appendix \ref{vortex}, we discuss the 
highly restrictive, non-local structure  of SU(N) duality transformations. We explicitly show the
non-trivial cancellations of infinite number of 
terms required to solve 
the SU(N) Gauss law constraints by the dual SU(N) 
spin 
operators. In  part 2 of Appendix \ref{vortex},  another set of  non-trivial 
cancellations are shown to hold for the SU(N) disorder operator to have a local physical action 
in the original Kogut-Susskind formulation. 
{\it In the case of much simpler $Z_2$ Wegner duality 
such cancellations are obvious.}
In Appendix \ref{appb}, we show that the `single spin' variational state satisfies Wilson's area law. 
In Appendix \ref{appc}, the expectation value of the truncated  dual spin Hamiltonian is computed in the variational ground state. The 
expectation value  of the non-local part of the spin Hamiltonian in the above disordered variational ground state is shown to vanish. This shows that  non-local terms 
appearing with higher powers of coupling $
(g^n, n \ge 3)$ may be treated perturbatively 
as $(g^2 \rightarrow 0)$ for the continuum.

Throughout this work, we use Hamiltonian formulation 
of lattice gauge theories \cite{ks} with open boundary conditions.  We work in two space 
dimensions on a finite lattice $\Lambda$ with ${\cal N} \left(=({\sf N}+1)\times ({\sf N}+1)\right)$ sites, ~${\cal L} \left(= 2{\sf N}({\sf N}+1)\right)$  
links, ${\cal P} \left(={\sf N}^2\right)$ plaquettes satisfying:
${\cal L} ={\cal P} + \left({\cal N}-1\right)$. 
 A lattice site is denoted by $(\vec n)$ or 
 $(m,n)$  with $m,n=0,1,\cdots ,{\sf N}$. The links are denoted by $(l)$ or $(m,n;~\hat i)$ 
with $i=1,2$. The plaquettes are denoted by the co-ordinates of their upper right corner and  sometimes by $p,p'$ etc.. Any conjugate pair operator $P,X$ satisfying the corresponding conjugate canonical commutation relations 
will be denoted by $\{P;X\}$.   In Ising model, Ising gauge theory, they are the Pauli 
spin operators $\{\sigma_1;\sigma_3\}$, $\{\mu_1;\mu_3\}$. In SU(N) lattice gauge theory, they are the Kogut-Susskind electric fields, link operators 
$\{E^a; U_{\alpha\beta}\}$. Similarly, the canonically conjugate  SU(N) spin, 
 string operators in the 
dual spin models  
are defined in the text.

\section{Duality and Canonical Transformations}
\label{sgsdct}
\subsection{Kramers-Wannier duality}
\label{skwdt} 
\begin{figure}
   	\includegraphics[scale=1.1]{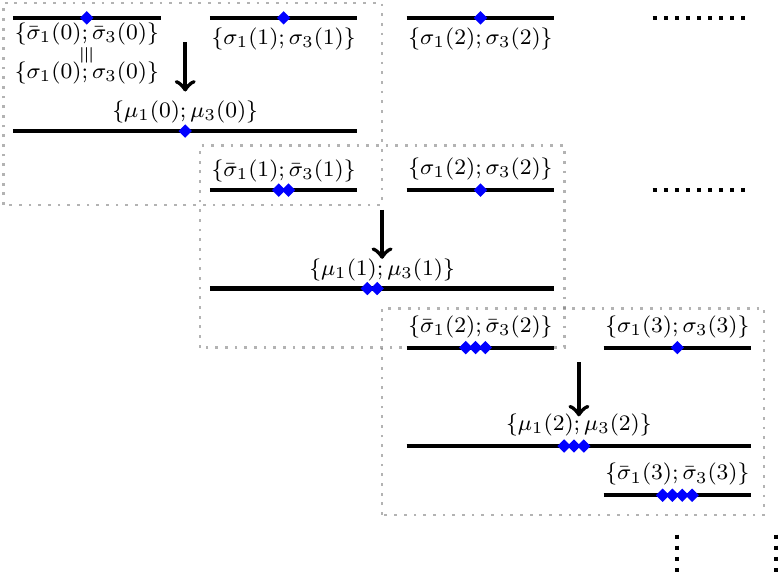}
   	\caption{Kramers-Wannier duality  through canonical transformations. The first three steps of duality or canonical  transformations in (\ref{ct1dim}) are illustrated.
   	}
   	\label{kramers}
   \end{figure}

Kramers Wannier duality was the first and the simplest duality, apart from electromagnetism, to be constructed. As a prelude to the construction of dualities in $Z_2$ and SU(N) lattice gauge theories, 
  we  apply canonical transformations  to $(1+1)$ dimensional Ising model to get the Kramers-Wannier duality. The Ising Hamiltonian in one space dimension 
  is in terms of the canonically conjugate operators $\{\sigma_1;\sigma_3\}$ at every lattice site 
  satisfying, 
  \begin{eqnarray}
  &\hspace{1.5cm}\sigma_1^2(m)=1;\hspace{1.5cm} \sigma_3^2(m)=1;  ~~~~~~~\label{sigmap} \\  
  &\sigma_1(m)\sigma_3(m)=-\sigma_3(m)\sigma_1(m)~~ {\textrm or} ~~[\sigma_1(m),\sigma_3(m)]_+ =0.\nonumber
  \end{eqnarray} 
    The Hamiltonian is
 \begin{align} 
 H ~=  ~\sum_{m=0}^{\infty} \Big[\sigma_1(m) -\lambda ~\sigma_{3}(m)\sigma_3(m+1)\Big].  
 \label{1dim} 
 \end{align} 
 The Kramers-Wannier duality is obtained by the 
 following iterative canonical transformations 
 along a line with $\bar \sigma_3(m=0) \equiv \sigma_3(m=0)$ and $\bar \sigma_1(m=0) \equiv \sigma_1(m=0)$: 
 \begin{eqnarray}
 &\mu_1(m)  \equiv \bar \sigma_3(m)\sigma_3(m+1), \nonumber\\ 
 & \mu_3(m) = \bar \sigma_1(m)   \label{ct1dim} \\ 
 &\bar \sigma_3(m+1) = \sigma_3(m+1), \nonumber\\
 &\bar \sigma_1(m+1) = \bar \sigma_1(m)\sigma_1(m+1)= \mu_3(m)\sigma_1(m+1). \nonumber
 \end{eqnarray} 
 The above canonical transformations iteratively replace the conjugate pair $\{\sigma_1(m);\sigma_3(m)\}$ or equivalently
 $\{\bar \sigma_1(m);\bar \sigma_3(m)\}$ 
 by a new conjugate pair
 $\{\mu_1(m);\mu_3(m)\}$. These new pairs are mutually independent and also satisfy the canonical relations (\ref{sigmap}). Unlike gauge theories (to be discussed in the next section), there are no spurious (string) degrees of freedom. This process    is graphically illustrated in Figure \ref{kramers}. 
 The relations (\ref{ct1dim}) lead to, 
 \begin{eqnarray} 
 \mu_3(m) = \prod_{s=0}^{m}\sigma_1(s).
 \label{mu1r} 
 \end{eqnarray} 
  The relations (\ref{mu1r}) can be easily inverted
  to give  
 $\sigma_1(m) = \mu_3(m)\mu_3(m-1)$ 
 with the convention $\mu_3(m=-1) \equiv 1$. The Ising model Hamiltonian can now be rewritten in its self-dual form in terms of the new dual conjugate pairs $\{\mu_1(m);\mu_3(m)\}$: 
 \begin{align} 
 H ~=  ~\sum_{m=0}^{\infty} \Big[\mu_3(m)\mu_3(m+1) -\lambda ~\mu_1(m)\Big].  
 \label{1dim} 
 \end{align} 
 Therefore, 
 \begin{align}          
 H(\sigma;\lambda)=\lambda^{-1} H(\mu;\lambda^{-1}).\nonumber
 \end{align}
 This is the famous Kramers-Wannier self duality.   As 
 expected, duality  has interchanged the interacting and non interacting parts of the Hamiltonian on going from the $\{\sigma_1,\sigma_3\}$ to the dual $\{\mu_1,\mu_3\}$ variables. In other words, 
 duality/canonical transformations (\ref{ct1dim}) map strong coupling region to the weak coupling region and vice versa. 
 
 \subsubsection{\bf Ising disorder operator}
   
    In Ising model  the magnetization operator, 
   $\sigma_3(m)$ is the order operator  as its expectation 
   value  measures the degree of order of the $\sigma_3$ variables. It is zero for $\lambda<\lambda_c$ and non-zero for $\lambda>\lambda_c$. This implies that the $\lambda>\lambda_c$ phase spontaneously breaks the  global $Z_2$ symmetry:  $\sigma_3 \rightarrow - \sigma_3$. On the other hand, the dual Hamiltonian 
   (\ref{1dim}) implies  that it is natural to define $\mu_3(m)$ as a disorder operator \cite{kogrev,fradsuss,horn}.
    The vacuum expectation value  ${}_{\lambda}\langle0|\mu_3(m)|0\rangle{}_{\lambda}$ is the disorder parameter. We also note that  the disorder operator $\mu_3({x_0})$
    acting on a completely ordered state (all $\sigma_3(m) =+1$ or 
   $-1$), flips all $\sigma_3$ spins at $m<{x}_0$ and creates a kink at $x_0$.  
   The resulting kink state is orthogonal to the original ordered state and the expectation value of the disorder operator $\mu_3$ in an ordered state vanishes: 
   \begin{align}
   {}_{{}_{\lambda=\infty}}\langle 0|\mu_3(m)|0\rangle_{{}_{\lambda=\infty}} =0, \hspace{.5cm}  {}_{{}_{\lambda=\infty}}\langle 0|\sigma_3(m)|0\rangle_{{}_{\lambda=\infty}} =1.  
   \label{smdr}
   \end{align} 
   \begin{figure}
       	\centering
       	\includegraphics[scale=0.75]{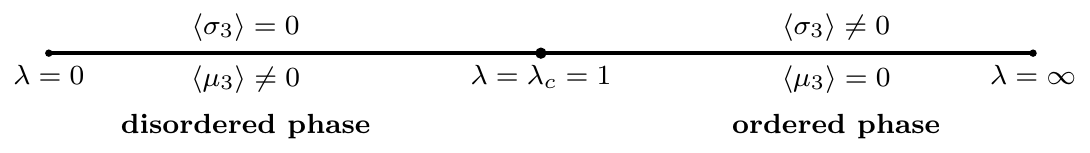}
       	\caption{Duality and  ordered \& disordered phases of $(1+1)$ dimensional  Ising model \cite{kogrev,fradsuss,horn}.} 
       	\label{orderdisorder} 
       \end{figure}
   \label{sz2gtsm}
   \begin{figure*}
      \centering
      \includegraphics[scale=0.8]{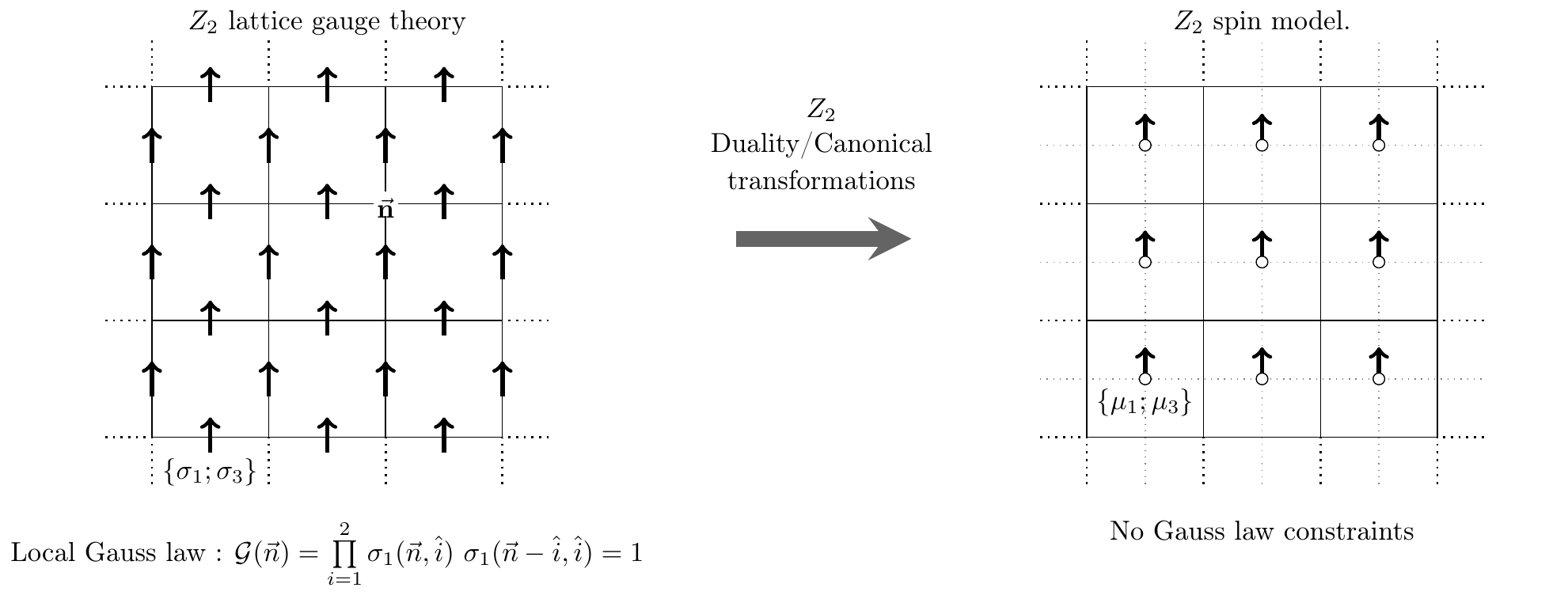}
       \caption{Duality between $Z_2$ lattice gauge theory and $Z_2$ (Ising) spin model.
       The initial and the final conjugate pairs 
       $\{\sigma_1;\sigma_3\}$ and $\{\mu_1;\mu_{3}\}$,  are  defined on the links and the plaquettes or dual sites respectively. The corresponding SU(N) 
       duality is  illustrated in Figure \ref{sundf}.} 
           \label{nglc}
      \end{figure*}
   
   On the other hand,  at $\lambda =0$,  the dual description  (\ref{1dim}) implies that the Ising model is in 
   ordered state with respect to  $\mu_3$. 
   As a consequence, 
   the disorder parameter does not vanish and order 
   parameter vanishes:  
   \begin{eqnarray}
   {}_{{}_{\lambda=0}}\langle 0|\mu_3(m)|0\rangle_{{}_{\lambda=0}} = 1, \hspace{1cm}  {}_{{}_{\lambda=0}}\langle 0|\sigma_3(m)|0\rangle_{{}_{\lambda=0}} =0.  
   \label{msdr}
   \end{eqnarray}
    The relations  (\ref{smdr}), (\ref{msdr}) are illustrated in Figure \ref{orderdisorder}.

\subsection{Wegner duality and $Z_2$ Spin Model}

$Z_2$ and $Z_N$ lattice gauge theories are the simplest 
theories with gauge structure and  many rich features.  
Due to their enormous simplicity compared to 
$U(1)$ or non-abelian lattice gauge theories and the presence of a confining phase, they have  been used as a
 simple theoretical laboratory to test 
 various confinement ideas \cite{horn}. 
 They also provide an explicit  realization of the Wilson-'t Hoofts algebra of order and disorder operators  characterizing 
 different possible phases of the SU(N) gauge
 theories \cite{hooft,horn}. In 1964, Schultz, Mattis and Lieb showed that the two-dimensional
 $Z_2$ Ising model
 is equivalent to a system of locally coupled fermions \cite{schultz}.  This result was later extended to 
 $Z_2$ lattice gauge theory which also  allows an equivalent description 
 in terms of  locally interacting fermions \cite{fradkinf}.  These are old and well known results. 
 In the recent past,   $Z_2$ $(Z_N)$ lattice gauge theories have been useful to understand quantum spin models \cite{fradkinb}, quantum computations \cite{kitaev}, tensor network or matrix product states  \cite{taka} and  their topological properties \cite{wen}, cold atom simulations \cite{reznik} and  entanglement entropy \cite{polik}.
In view of these wide applications,  the 
$Z_2$ $(Z_N)$ lattice gauge theories and 
associated duality transformations are important in their 
own right. 
 
The $Z_2$ lattice gauge theory involves $Z_2$ conjugate spin operators $\{\sigma_1(l); \sigma_3(l)\}$  on the  link $l \in \Lambda$. The anti-commutation relations amongst 
these conjugate pairs on every link $l$ are 
\begin{align} 
\sigma_1(l) ~\sigma_3(l)+\sigma_3(l) ~\sigma_1(l)=0. 
\label{ccrz2} 
\end{align} 
They further satisfy: ${\sigma}_3(l)^2={\sigma}_1(l)^2=1$. 
In order to  maintain a 1-1 correspondence with SU(N) lattice gauge theory (discussed in the next section), it is convenient to identify the conjugate pairs $\{\sigma_1(l); \sigma_3(l)\}$ 
with $Z_2$ electric field, $E(l)$ and 
 $Z_2$  vector potential, $A(l)$ as: \begin{eqnarray} 
 \sigma_1(l)= e^{i\pi E(l)}, ~~~~~~ \sigma_3(l) = e^{iA(l)}.
 \label{z2efmp} 
 \end{eqnarray}  
Above  $E(l) \equiv \{0, 1\}$ and $ A(l) \equiv \{0, \pi\}$.
A basis of the two dimensional Hilbert space  on each link $l$ is chosen to be  the eigenstates $\ket{\pm , l}$ of $\sigma_3(l)$ with eigenvalue $ \pm 1$ with 
$\sigma_1(l)$ acting as a spin flip operator:
\begin{align} 
\sigma_3(l) \ket{\pm,l} = \pm \ket{\pm,l}, ~~ ~~~\sigma_1(l) |\pm ,l\rangle = |\mp ,l\rangle.
\end{align}
  
The $Z_2$ lattice gauge theory Hamiltonian is given by
 \begin{align}
 H&=-\sum_{l \in \Lambda} \sigma_1(l)- \lambda \sum\limits_{p \in \Lambda} \sigma_3(l_1) \sigma_3(l_2) \sigma_3(l_3) \sigma_3(l_4) \nonumber \\ 
 &\equiv H_E +\lambda H_B.
 \label{isingH}
 \end{align}
 In (\ref{isingH}) $\sigma_3(l_1) \sigma_3(l_2) \sigma_3(l_3) \sigma_3(l_4)$ represents the product of $\sigma_3$ operators along the four links of a  plaquette. The sum over $l$ and $p$ in (\ref{isingH}) are the sums over all  links and plaquettes
 respectively. The 
 parameter $\lambda$ is the $Z_2$ gauge theory coupling constant.
  The first term $H_E$  and the second term $H_B$ in (\ref{isingH}) represent  the $Z_2$ electric and 
 magnetic field operators respectively. 
 The electric field operator $\sigma_1(l)$ is fundamental while the  latter  
  is a composite of the four $Z_2$ magnetic 
  vector potential operators  $\sigma_3(l)$ 
  along a plaquette.  After a series of canonical transformations, the above  characterization 
  of electric, magnetic field   will be reversed. More explicitly,  the  dynamics will be described by  the Hamiltonian (\ref{isingH}) rewritten in terms of the fundamental  magnetic field (the second term) and the  electric field operator (the first term) 
  will be composite of the dual electric scalar potentials (see (\ref{plaqop1}) and (\ref{iroel2})). The same feature will be repeated in the SU(N) case discussed in the next section. 

\begin{figure}
  	\centering
  	\includegraphics[scale=.9]{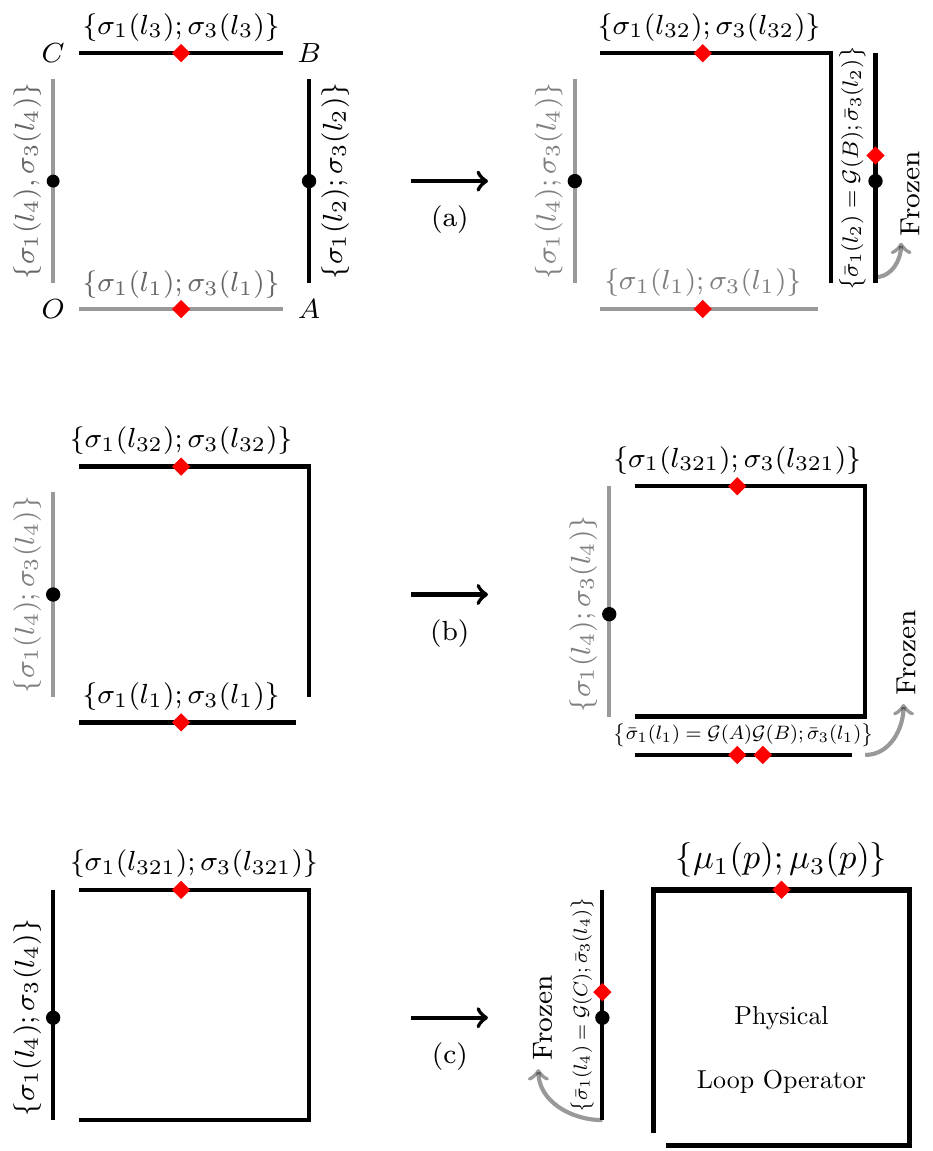}
  	\caption{The $Z_2$ canonical transformations (\ref{zct1}), (\ref{zct2}), (\ref{zct3a}) and (\ref{zct3b}) are pictorially illustrated  in (a), (b) and (c) respectively. The ${\color{red} \blacklozenge}$ and $\bullet$ represent the electric fields of  the initial  horizontal and vertical links respectively.}  
  	\label{plaqct}
  \end{figure}

 The Hamiltonian (\ref{isingH}) remains invariant if all 4 spins attached  to the 4 links emanating from a site $n$ 
 are flipped simultaneously. 
 This symmetry operation is implemented by  the Gauss law operator ${\cal G}$: 
 \begin{align}  
  {\cal G}(n)\equiv\prod_{l_n} \sigma_1(l_n) 
  \label{z2gl}
  \end{align}
  at lattice site $n \in \Lambda$. 
  In (\ref{z2gl}), $\prod_{l_n}$ represents the product over 4 links (denoted by $l_n$) which share the lattice site 
  $n$ in two space dimensions. The  $Z_2$ gauge transformations are 
 \begin{align}
 \sigma_1(l) &\rightarrow {\cal G}^{-1}(n){\sigma}_1(l) {\cal G}(n)={\sigma}_1(l), ~~~~~\forall ~l \in \Lambda, \nonumber \\
  \sigma_3(l_n) &\rightarrow {\cal G}^{-1}(n){\sigma}_3(l_n) {\cal G}(n)=-{\sigma}_3(l_n),\\ H &\rightarrow {\cal G}^{-1}(n)~H~{\cal G}(n)=H. \nonumber
 \end{align} 
Thus, under a gauge transformation at site $n$, the 4 link flux operator $\sigma_3(l_n)$ on the  4 links $l_n$ sharing the lattice site $n$ change sign.
 All other $\sigma_3(l)$  remain invariant.  
 The physical Hilbert space ${\cal H}^p$ consists 
 of the states $\ket{phys}$ satisfying the Gauss law constraints: 
  \begin{align} 
 {\cal G}(n) \ket{phys} & = \ket{phys}& {\textrm{or}} && {\cal G}(n) \approx 1 && \forall n \in \Lambda. 
 \label{zglc} 
 \end{align}
 In other words, ${\cal G}(n)$  are  unit 
 operators  within the physical Hilbert space 
 ${\cal H}^p$. All operator identities valid only within ${\cal H}^p$ are  expressed by $\approx$ sign.
     We now canonically transform this simplest $Z_2$ gauge theory with constraints (\ref{z2gl}) at every lattice site  into $Z_2$ spin model  without any constraints as shown in Figure \ref{nglc}. 
To keep the discussion simple,  we start with a single plaquette OABC  shown in Fig. \ref{plaqct}-a before 
dealing with the entire lattice. As the canonical transformations are iterative in nature, this simple example contains all the essential ingredients required to understand the finite lattice case. The four  links 
OA, AB, BC, CO will be denoted by $l_1,l_2,l_3, l_4$ 
respectively. In this simplest case there are four  $Z_2$ gauge transformation or equivalently Gauss law operators (\ref{z2gl}) at each of the four corners O, A, B and  C:
\begin{align}
{\cal G}(O) &= {\cal G}(0,0) = \sigma_1(l_4)\sigma_1(l_1) \approx 1, \nonumber\\ 
{\cal G}(A) &= {\cal G}(1,0)= \sigma_1(l_1)\sigma_1(l_2) \approx 1,   \nonumber \\ 
{\cal G}(B) &= {\cal G}(1,1) = \sigma_1(l_2)\sigma_1(l_3) \approx 1,\nonumber\\
 {\cal G}(C) &= {\cal G}(0,1) = \sigma_1(l_3)\sigma_1(l_4) \approx 1.
\label{glp} 
\end{align}  
Note that these Gauss law operators satisfy a trivial operator identity:
\begin{align} 
{\cal G}(O)~{\cal G}(A)~{\cal G}(B)~{\cal G}(C) ~\equiv ~1.
\label{ggc} 
\end{align}  
The above identity states the obvious result  that a simultaneous  flippings   at all 4 sites has no effect. This is because of the abelian nature of the 
gauge group.
We now  start with the four  initial conjugate pairs on links $l_1,l_2,l_3$ and $l_4$: \begin{align}
\{\sigma_1(l_1);\sigma_3(l_1)\}&,&  
\{\sigma_1(l_2);\sigma_3(l_2)\}, \nonumber\\
 \{\sigma_1(l_3);\sigma_3(l_3)\}&,& \{\sigma_1(l_4);\sigma_3(l_4)\}.
\end{align}
Using canonical transformations we  define four 
new but equivalent conjugate  pairs. The first three string conjugate pairs:   
$$\{\bar \sigma_1(l_1);\bar \sigma_3(l_1)\},~~  
\{\bar \sigma_1(l_{2});\bar \sigma_3(l_{2})\},~~ \{\bar \sigma_1(l_4); \bar \sigma_3(l_4)\}$$ 
describe the  collective excitations on the links $OA,~AB,~ BC$  and shown in Figures \ref{plaqct}-b,a,c 
respectively. The remaining collective excitations over the plaquette or the loop $p\equiv OABC$ are described by $$\{\mu_1(p);\mu_{3}(p)\}$$ and shown in Figure \ref{plaqct}-c.
 As a consequence of the 
 three mutually independent Gauss law constraints ${\cal G}(A), {\cal G}(B)$ and ${\cal G}(C)$, the three string electric fields are frozen to the value $+ 1$.  Therefore there is no dynamics associated with the
 three  strings.
 In other words, string degrees of freedom completely decouple from  
${\cal H}^{p}$.  We are thus left with the 
final physical  $Z_2$ 
spin operators  $\{\mu_1(p);\mu_{3}(p)\}$ which are 
explicitly $Z_2$ gauge invariant. These duality 
transformations from gauge variant link operators to gauge invariant spin or loop operators are shown in Figure \ref{nglc}.   To demonstrate the
above  results, we start with  the initial link  operators   $\{\sigma_1(l_3);\sigma_3(l_3)\}$ and 
$\{\sigma_1(l_2);\sigma_3(l_2)\}$ as 
shown in Fig. (\ref{plaqct})-a.
As was done in $(1+1)$ dimensional Ising model, 
we glue them using canonical transformations as 
follows:
    \begin{align}
 \bar\sigma_{3}(l_2) &\equiv ~~~\sigma_3(l_2), &\sigma_{3}(l_{32}) &\equiv \sigma_3(l_3) \sigma_3(l_2)\nonumber \\
 \bar\sigma_{1}(l_2)&=\sigma_1(l_3) \sigma_1(l_2) \equiv {\cal G}(B),&\sigma_{1}(l_{32}) &= \sigma_1(l_3).
 \label{zct1}
 \end{align}
  The canonical transformations (\ref{zct1}) are  illustrated in Fig. \ref{plaqct}-a. After the transformations,  
  the two new but equivalent canonical sets  $\{\bar\sigma_1(l_2)={\cal G}(B); \bar\sigma_3(l_2)\}$,  $\{\sigma_1(l_{32}); \sigma_3(l_{32})\}$  are attached to the links $l_2$ and  $l_{32}\equiv l_3l_2 $ respectively. They  satisfy the same commutation relations as the original operators (\ref{ccrz2}): 
\begin{align}
 \bar\sigma_1(l_2)\bar\sigma_3(l_2) &+\bar\sigma_3(l_2)\bar\sigma_1(l_2)=0,\\   
 \sigma_1(l_{32})\sigma_3(l_{32}) &+ \sigma_3(l_{32})\sigma_1(l_{32}) =0.  \nonumber
 \end{align}
One can easily check: $\bar \sigma_1^2(l_2)=1,~\bar\sigma_3^2(l_2) =1,  
\sigma_1(l_{32})^2=1,\sigma_3(l_{32})^2 =1.$
Further, note that the two  conjugate pairs $\{\bar\sigma_1(l_2);\bar\sigma_3(l_2)\}$ 
and $\{\sigma_1(l_3l_2); \sigma_3(l_3l_2)\}$  are also mutually independent as they commute with each other. As an example,  $\left[\bar\sigma_1(l_2),\sigma_3(l_3l_2)\right] \equiv
\left[\sigma_1(l_3) \sigma_1(l_2),\sigma_3(l_3) \sigma_3(l_2)\right] =0$.  
The new conjugate pair $\{\bar \sigma_1(l_2); \bar \sigma_3(l_2)\}$ is frozen due to the Gauss law at B: 
$\bar{\sigma}_1(l_2)={\cal G}(B) \approx 1$ in ${\cal H}^p$. 
We now repeat ({\ref{zct1}) with $l_2,l_3$ replaced by $l_{1},l_{32}$ respectively 
to define new conjugate operators 
$\{\bar \sigma_1(l_1);\bar \sigma_3(l_1)\}$ 
and $\{\sigma_1(l_{321});\sigma_3(l_{321})\}$ attached 
to the links $l_1$ and $l_{321} (\equiv l_3l_2l_1)$ respectively:
 \begin{align}
 \label{zct2}
 &\bar\sigma_{3}(l_1) \equiv \sigma_3(l_1), ~~~~ ~~~~~~~
 \sigma_{3}(l_{321}) \equiv \sigma_3(l_{32}) \sigma_3(l_1) \\
 &~ \bar\sigma_{1}(l_1)=\sigma_1(l_{32}) \sigma_1(l_1)
   ={\cal G}(A){\cal G}(B)
 ,~~ \sigma_{1}(l_{321}) = \sigma_1(l_3). \nonumber  
\end{align}
 As before, the new conjugate pair $\{\bar{\sigma}_1(l_1); \bar{\sigma}_3(l_2) \}$ becomes unphysical as $\bar{\sigma}_1(l_1)={\cal G}(A){\cal G}(B)\approx 1$ in ${\cal H}^p$.
The last canonical transformations involve 
gluing the conjugate pairs $\{\sigma_1(l_{321}); \sigma_3(l_{321})\}$ with 
$\{\sigma_1(l_{4});\sigma_3(l_{4})\}$ to define the dual and gauge invariant plaquette variables $\{\mu_1(p); \mu_{3}(p)\}$, 
with $p\equiv l_1l_2l_3l_4$:
\begin{subequations}
\begin{align}
 &   \mu_{1}(p) \equiv \sigma_{3}(l_{321})\sigma_{3}(l_4) \equiv \sigma_3(l_{3}) \sigma_3(l_{2})\sigma_3(l_1)\sigma_3(l_4).  \nonumber \\
 & ~~~~~~~~~~~~~~\mu_3(p) \equiv \sigma_{1}(l_{321}) ~= \sigma_1(l_{3}), 
  \label{zct3a}
\end{align} 
\begin{align}
\label{zct3b}
&~~~~~~~~~~~~~~\bar\sigma_{3}(l_{4}) \equiv \sigma_3(l_{4}),\\
&\bar\sigma_{1}(l_{4})=\sigma_1(l_{321}) \sigma_1(l_{4}) = \sigma_1(l_3)\sigma_1(l_4) \equiv {\cal G}(C). \nonumber 
 \end{align} 
 \end{subequations}
To summarize, the three canonical transformations (\ref{zct1}), (\ref{zct2}),  (\ref{zct3a}) and (\ref{zct3b}) transform 
the initial four conjugate sets   $\{\sigma_1(l_1);\sigma_3(l_1)\}$,$ 
\{\sigma_1(l_2);\sigma_3(l_2)\}$,$\{\sigma_1(l_3);\sigma_3(l_3)\}$, $~\{\sigma_1(l_4);\sigma_3(l_4)\}$ attached to the links $l_1,l_2,l_3,l_4$ to four new and equivalent canonical sets 
$\{\bar \sigma_1(l_{2});\bar \sigma_3(l_{2})\},~\{\bar \sigma_1(l_1);\bar \sigma_3(l_1)\},~
\{\bar \sigma_1(l_4);\bar \sigma_3(l_4)\}$ and $\{\mu_1(p);\mu_{3}(p)\}$ attached to the links 
$l_2,l_{1},l_4$ and the plaquette $p$ respectively.
The advantage of the new sets is that 
all the three independent Gauss law constraints  
at $A,B$ and $C$
are automatically solved. They freeze  
the 
three strings leaving us  only with  the physical spin or plaquette  loop conjugate operators $\{\mu_1(p);\mu_{3}(p)\}$. The defining canonical relations (\ref{zct1}), (\ref{zct2}), (\ref{zct3a}) 
and (\ref{zct3b}) can also be inverted. The inverse transformations from  the new spin flux operators to $Z_2$ link flux operators are
\begin{eqnarray} 
\label{loit} 
\sigma_3(l_1) = \bar \sigma_3(l_1), ~~~~\sigma_3(l_2) = \bar \sigma_3(l_{2}),~~~~~~~~~~~~\\ 
\sigma_3(l_3)= \mu_{1}(p) \bar \sigma_3(l_{4})\bar \sigma_3(l_1)\bar \sigma_3(l_2), ~~~~\sigma_3(l_4) =\bar \sigma_3(l_4). \nonumber
\end{eqnarray}  
Similarly, the initial conjugate $Z_2$ electric field operators on the links are
\begin{align} 
\label{oefa}
\sigma_1(l_1) & = \mu_3(p)~ \bar \sigma_1(l_1)= \mu_3(p) ~{\cal G}(A){\cal G}(B) \approx \mu_3(p), \nonumber \\
\sigma_1(l_2) & = \mu_3(p) ~\bar \sigma_1(l_{2}) = \mu_3(p) ~{\cal G}(B) \approx \mu_3(p) \\
\sigma_1(l_3) & = \mu_3(p), \nonumber\\ \sigma_1(l_4) &= \mu_3(p) ~\bar \sigma_1(l_4) = \mu_3(p) ~{\cal G}(C) \approx \mu_3(p).  \nonumber 
\end{align}
Thus the complete set of gauge-spin duality relations over a plaquette and their inverses are  given in (\ref{zct1}), (\ref{zct2}), (\ref{zct3a}), (\ref{zct3b}) and  (\ref{loit}), (\ref{oefa}) respectively.  
Note that the Gauss law constraint at the origin does not play any role as ${\cal G}(O) \approx {\cal G}(A)~{\cal G}(B)~{\cal G}(C)$. The total number of degrees of freedom also match. The initial $Z_2$ gauge theory 
had 4 spins with 3 Gauss law constraints. In the final dual spin model the 3 gauge non-invariant strings take care of the 3 Gauss law constraints leaving us  with the single gauge invariant  
spin described by $\{\mu_1(p);\mu_{3}(p)\}$ on the plaquette $p$. 
The single plaquette $Z_2$ lattice gauge theory Hamiltonian (\ref{isingH}) can now be  rewritten in terms of the new gauge invariant spins as: 
   \begin{align} 
 H \approx - 4 ~\mu_3(p) -\lambda ~\mu_{1}(p)  
 = - \left(\begin{array}{ccc}  \lambda ~ & 4   \\ 
 4 ~& - \lambda   
\  \end{array} \right).
 \label{isingloopH} 
 \end{align}
Note that the equivalence of the gauge and spin Hamiltonians (\ref{isingH}) and (\ref{isingloopH}) respectively is valid only within the physical Hilbert space ${\cal H}^p$. 
The two energy eigenvalues of $H$ are $\epsilon_\pm = 
 \pm  4 \sqrt{ \left(1 + \left(\frac{\lambda}{4}\right)^2\right)}$. 

 Having discussed the essential ideas, we now directly write down  the general $Z_2$ gauge-spin duality or canonical relations 
 over the entire lattice.  The details of these 
 iterative canonical transformations (analogous to (\ref{zct1}), (\ref{zct2}), (\ref{zct3a}) 
 and (\ref{zct3b})) are given in Appendix \ref{appa}.  Note that 
 there are ${\cal L}$ initial spins (one on  every link) with  ${\cal N}$ Gauss law constraints (one at every site) satisfying the identity: 
 \begin{align} 
 \prod_{(m,n) \in \Lambda} ~{\cal G}(m,n) ~\equiv ~1. 
 \label{z2gli}
 \end{align}
 The above identity again  states that 
 simultaneous flipping of all spins around every lattice site is an identity operator because each spin is flipped twice. As mentioned earlier, it is a property of all abelian gauge theories  which reduces the number 
    of Gauss law constraints from ${\cal N}$ to $({\cal N}-1)$.  In the non-abelian SU(N) case, discussed in the next section, there is no such reduction. 
    The global SU(N) gauge transformations, corresponding to the extra Gauss law constraints at the origin ${\cal G}^a(0,0) =1$, need to be fixed  by hand to get the correct number of physical degrees of freedom (see section \ref{rgl}). 
    After canonical transformations in $Z_2$ lattice gauge theory,  there are 
    (a) ${\cal P}$ physical plaquette spins (analogous to $\{\mu_1(p);\mu_{3}(p)\}$ in the single plaquette case) shown in Figure \ref{dual}-a  and (b) $({\cal N}-1)$ stringy 
 spins (analogous to $\{\bar \sigma_1(l_1);\bar \sigma_3(l_1)\}; ~\{\bar \sigma_1(l_2);\bar \sigma_3(l_2)\}$ and $\{\bar \sigma_1(l_{4});\bar \sigma_3(l_{4})\}$ in the single plaquette case)
as every lattice site away from the origin can be attached to a unique string. This is shown in Figure \ref{dual}-b.
The degrees of freedom before and after the canonical transformations match as ${\cal L} = {\cal P} + ({\cal N} -1)$.    
All  $({\cal N} -1)$ strings decouple because of the  $({\cal N}-1)$ Gauss law constraints. The algebraic details of these  transformations leading to  freezing of all strings are 
 worked out in detail in Appendix \ref{appa}. 
 
 From now onward the  ${\cal P}$ physical plaquette 
    spin/loop operators are labelled by the top right corners of the corresponding plaquettes as shown in Figure \ref{dual}-a).  
  The vertical (horizontal) stringy spin operators are labelled by the top (right) end points of the corresponding links as shown in  
  Figure \ref{dual}-b. The same notation will be used to label the dual SU(N) operators in section \ref{snalgt}.
 \begin{figure}
    \centering
     \includegraphics[scale=0.6]{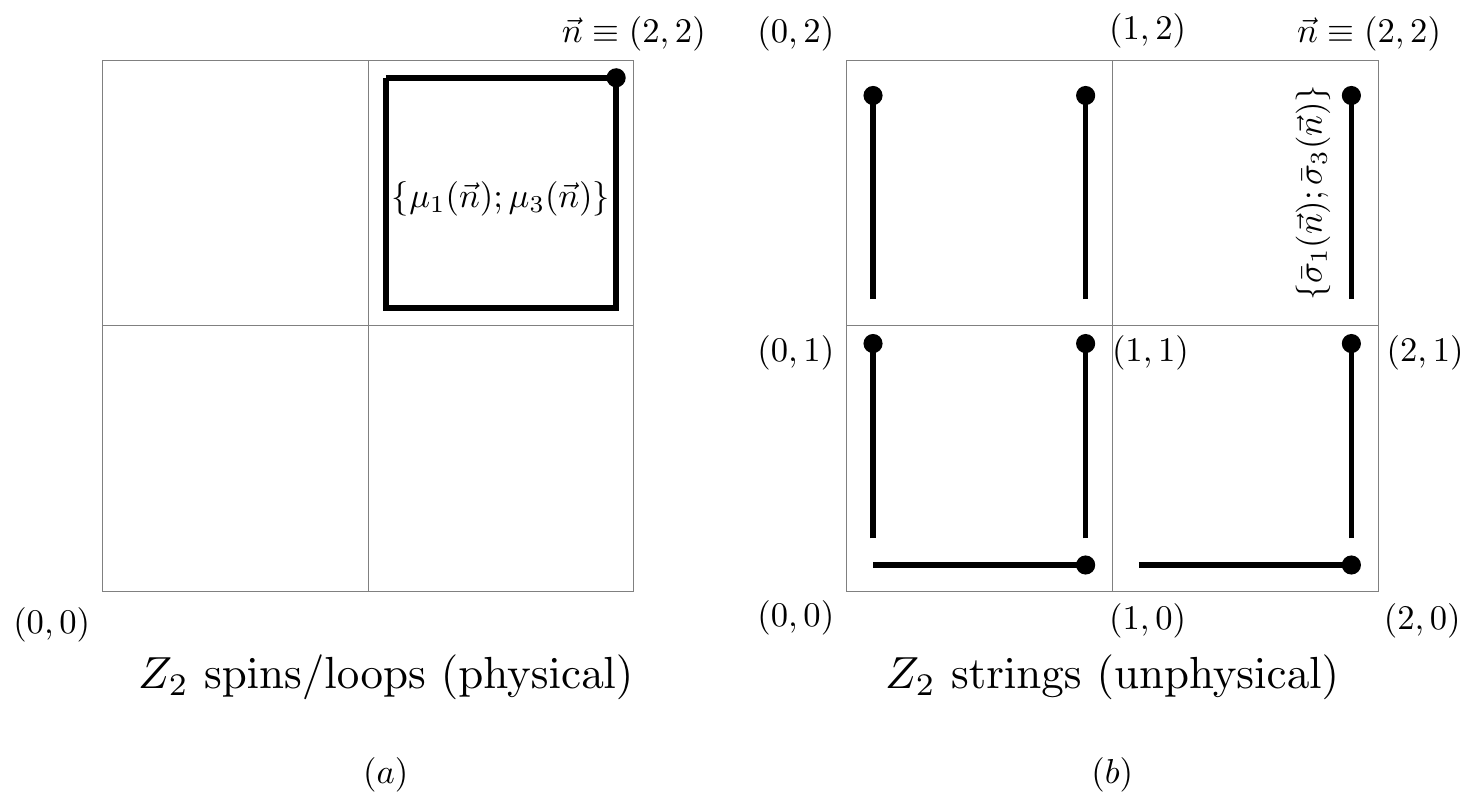}
     \caption{The physical $Z_2$ spin conjugate pairs $\{\mu_1(\vec n); \mu_{3}(\vec n)\}$ and the unphysical string conjugate pairs $\{\bar{\sigma}_1(\vec n); \bar{\sigma}_3(\vec n)\}$ dual to $Z_2$ lattice gauge theory are shown in (a) and (b) respectively. 
     The co-ordinates of spin or loop operators are 
     the co-ordinates of their top right corners. 
     The co-ordinates of the horizontal (vertical) 
     strings are the co-ordinates of their right (top) 
     end points. These are shown by $\bullet$ in (a) and (b). 
      The strings decouple from the physical Hilbert space as $\bar{\sigma}_1(\vec n) = {\cal G}(\vec{n}) \approx 1$ by Gauss law  
     constraint at $\vec  n$. The corresponding dual SU(N) spin and SU(N) string operators are shown in Figure \ref{dualsun}-a,b respectively.}
     \label{dual}
    \end{figure} 
    \begin{figure*}
               \includegraphics{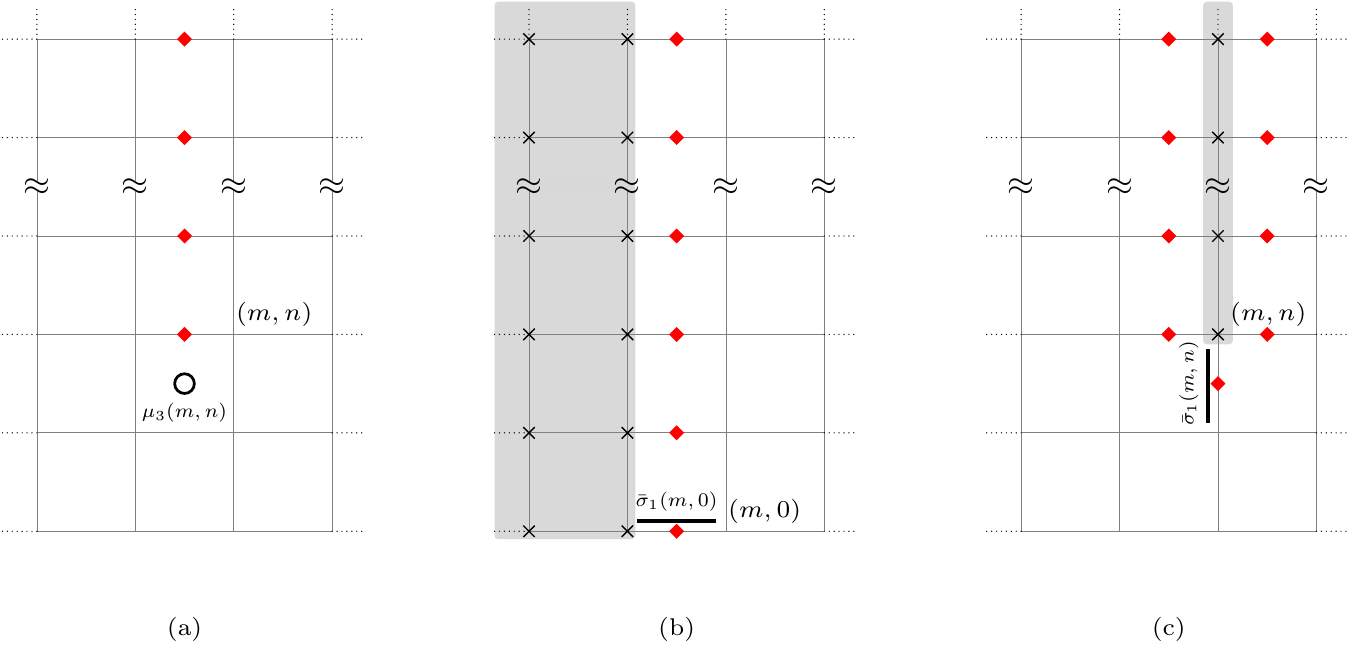}
               \caption{ The non-local relations in the $Z_2$ gauge-spin duality transformations: 
                (a) shows the relations (\ref{plaqop2}) expressing  $\mu_3(m,n)$ as the product of $\sigma_1$ operators denoted by {\tiny{\color{red}$\blacklozenge$}}.  In (b) and (c), we show the relations 
                 (\ref{z2duality2}) expressing  $\bar{\sigma}_1(m,0)$ and $\bar{\sigma}_1(m,n); ~n\neq0$ respectively as the product of $\sigma_1$ operators denoted by {\tiny{\color{red}$\blacklozenge$}}. As $\sigma_1^2=1$, the string operators $\bar \sigma_1(m,0)$ and $\bar \sigma_1(m,n)$ are also a product of Gauss law operators at sites marked by x in the shaded regions. For similar SU(N) relations, see Figures \ref{frelsun}. }
               \label{rellz2} 
               \end{figure*} 
    \subsubsection{\bf  Physical sector and $Z_2$ dual potentials}   
    The final duality relations between  the initial conjugate sets  $\{\sigma_1(m,n;\hat{i});\sigma_3(m,n; \hat{i})\}$ on every lattice link $(m,n; \hat i)$ and the final physical conjugate loop operators 
 $\{\mu_1(m,n)$; $\mu_{3}(m,n)\}$
 are (see Appendix \ref{appa})
 \begin{subequations}
 \begin{align}
 \mu_{1}(m,n)&=\sigma_3(m-1,n-1;\hat{1})~\sigma_3(m-1,n-1;\hat{2})\nonumber \\
  &~~~~~~~~\sigma_3(m,n; -\hat{2}) \sigma_3(m,n;-\hat{1}),
 \label{plaqop1}
 \end{align}
 \begin{align}
  \mu_3(m,n)=\prod\limits_{n'= n}^{N} \sigma_1(m-1,n';\hat{1}). 
 \label{plaqop2} 
 \end{align}
 \end{subequations}
 In (\ref{plaqop1}) we have defined $\sigma_1(m,n;-\hat 1) \equiv \sigma_1(m-1,n; \hat 1)$ and $\sigma_1(m,n;-\hat 2) \equiv \sigma_1(m,n-1; \hat 2)$.
 The relations (\ref{plaqop1}) and (\ref{plaqop2})  are the extension of the  single plaquette relations (\ref{zct3a}) to the entire lattice. They are illustrated in Figure \ref{rellz2}-a.  The canonical commutation relations are 
 \begin{align} 
 \mu_1(m,n)\mu_{3}(m,n)+\mu_{3}(m,n)\mu_1(m,n) = 0.  
 \end{align} 
  Further, $\mu_3(m,n)^2 =1, \mu_{1}(m,n)^2=1$. 
 {\it The canonical transformations  (\ref{plaqop1}) are important as they define the  magnetic field operators $\mu_1(m,n)$ and its conjugate $\mu_3(m,n)$  
 as a new  dual fundamental operators. The electric field is derived from the electric scalar potentials.} This should be contrasted with the  original description 
 where  electric fields $\sigma_1(m,n)$ were fundamental and  
 the magnetic fields were derived from the magnetic 
  magnetic vector potentials as  $\sigma_3(l_1)\sigma_3(l_2)\sigma_3(l_3)
 \sigma_3(l_4)$. 
 \subsubsection{\bf Unphysical sector and $Z_2$ string operators}    
 The unphysical string conjugate pair operators are
 (see Appendix \ref{appa}) 
  \begin{subequations}
 \begin{align}
 \bar{\sigma}_3(m,0)&=\sigma_3(m-1,0;\hat{1}),\nonumber\\
   \bar{\sigma}_3(m,n)&= \sigma_3(m,n-1;\hat{2});~~~~n\neq 0 
    \label{z2duality1}
   \end{align}
 
   \vspace{-1.cm}
 
 \begin{align} 
  \bar{\sigma}_1(m,0)&=\prod\limits_{m'=0}^{m-1} \prod\limits_{n'=0}^N {\cal G}(m',n') \approx 1,\nonumber\\
    \bar{\sigma}_1(m,n)
  &=\prod\limits_{n'=n}^N{\cal G}(m,n') \approx 1;~~~~n\neq 0. 
   \label{z2duality2}
 \end{align}
   \end{subequations}
%
 
  The relations (\ref{z2duality1}) and   (\ref{z2duality2}) are illustrated in Figure
  \ref{rellz2}-b  and Figure
    \ref{rellz2}-c  respectively. 
  It is easy to see that in the full gauge theory Hilbert space  $\bar \sigma_1(m,n) \bar \sigma_3(m,n) +  \bar \sigma_3(m,n) \bar \sigma_1(m,n) 
   =0$ and  different string operators located at different lattice sites commute with 
   each others.  
 Further, one can check that all strings and plaquette operators are mutually independent and commute with each other: 
  \begin{align} 
  \label{spcr}
 \left[\mu_3(m,n),\bar \sigma_1(m',n')\right] = 0, \left[\mu_3(m,n),\bar \sigma_3(m',n')\right] = 0,   \\
  \left[\mu_{1}(m,n),\bar \sigma_1(m',n')\right] = 0, \left[\mu_{1}(m,n),\bar \sigma_3(m',n')\right] = 0.
  \nonumber  
   \end{align}  
  \subsubsection{\bf Inverse relations}  
 The inverse relations for the flux operators over the entire lattice are
 \begin{align}
\sigma_3(m,0;\hat{1})&= \bar{\sigma}_3(m+1,0),\nonumber\\
 \sigma_3(m,n;\hat{2})&= \bar{\sigma}_3(m,n+1) \nonumber \\
\sigma_3(m,n;\hat{1})&= \bigg(\prod\limits_{l=1}^n\bar{\sigma}_3(m,l)\bigg)
\bigg(\prod\limits_{q=1}^n\bar{\sigma}_3(m+1,q)\bigg)
\nonumber\\
&~~~~~~~~~\bigg(\prod\limits_{p=1}^n{\mu}_1(m+1,p)\bigg);~~n \neq 0 
\label{iroel}
\end{align} 
On the other hand, the conjugate electric field operators are  
\begin{align} 
 \label{iroel2} 
 \sigma_1(m,n;\hat{1})= \mu_3(m,n)\mu_3(m,n+1), ~~~~~~\nonumber\\
 \sigma_1(m,n;\hat{2}) = \mu_3(m,n+1)\mu_3(m+1,n+1).   
\end{align} 
In the second relation in (\ref{iroel2}), we have used Gauss laws at $(m,l) ~; l=n+1,n+2, \cdots $.
The above relations are analogous to the inverse relations (\ref{loit}) and (\ref{oefa}) in the single plaquette case. 
 
 \subsubsection{\bf $Z_2$ Gauss laws \& solutions}
 \label{sz2glt}
 \begin{figure}[b]
 	\includegraphics[scale=0.9]{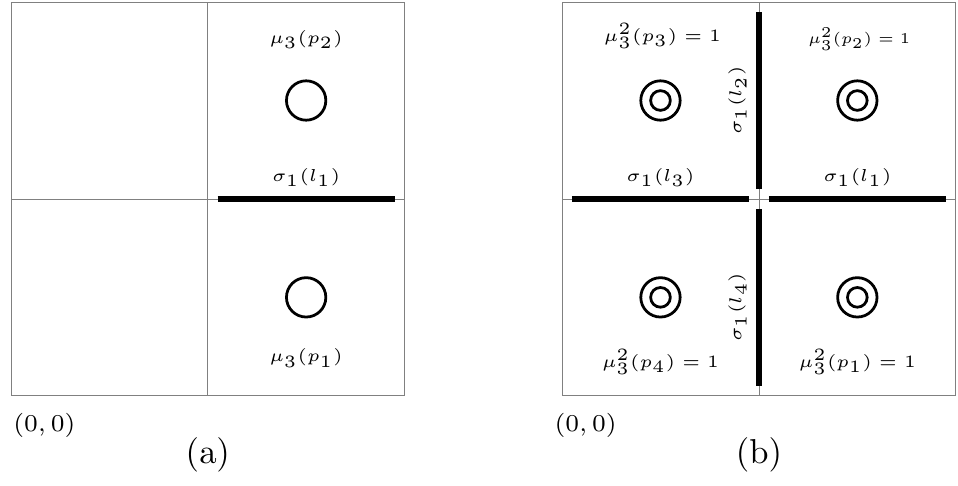}
 	\caption{(a) shows the $Z_2$ link electric field operator $\sigma_1(m,n;\hat{1})$ as the product of nearest neighbor loop operators $\mu_3(p_1)$ and $\mu_3(p_2)$, (b) graphically illustrates  how the spin or electric potential operators 
 	$\{\mu_3(p_1),\mu_3(p_2),\mu_3(p_3),\mu_3(p_4)\}$ solve the $Z_2$ Gauss law (\ref{z2glprfe}) at site $(m,n)$. A similar SU(N) proof is involved and given in Appendix \ref{nagls}.} 
 	\label{z2glc}
 \end{figure}
It is easy to see that the Gauss law constraints
are automatically satisfied by the dual spin operators
as shown in Figure \ref{z2glc}-a,b. 
We  write the $Z_2$ electric fields around a site $(m,n)$ in terms of the  electric scalar potentials: 

 \begin{align}
 & \sigma_1(l_1) \equiv \sigma_1(m,n;\hat{1})= \mu_3(p_1)\mu_3(p_2),
 \nonumber \\
 &\sigma_1(l_2) \equiv \sigma_1(m,n;\hat{2})= 
 \mu_3(p_2)\mu_3(p_3), \nonumber
 \end{align}
 \begin{align}
 &\sigma_1(l_3) \equiv \sigma_1(m-1,n;\hat{1})= 
 \mu_3(p_3)\mu_3(p_4), \nonumber \\
 &\sigma_1(l_4) \equiv \sigma_1(m,n-1;\hat{2})= 
 \mu_3(p_4)\mu_3(p_1).
 \label{tglc}
 \end{align}
In (\ref{tglc}) we have used link and plaquette labels from Figure \ref{z2glc}. As $\mu_3^2(p)=1$,  we get 
 \begin{align}
 {\cal G}(m,n)& = \sigma_1(l_1)\sigma_1(l_2)\sigma_1(l_3)\sigma_1(l_4)
 \equiv 1.
 \label{z2glprfe}
 \end{align}
 The above duality property also generalizes to the SU(N) case. {\it The dual SU(N) spin operators 
 or potentials are the solutions of local SU(N) Gauss laws at all the sites except origin}. However, unlike the trivial cancellations above, the non-abelian cancellations are highly nontrivial  and are worked out in detail in 
 Appendix \ref{nagls}. 
\subsubsection{\bf $Z_2$ dual dynamics} 
\label{z2dd} 

The $Z_2$ lattice gauge theory Hamiltonian (\ref{isingH}) in terms of the physical  spin 
operators takes the simple nearest neighbor interaction form:
\begin{align} 
 & H=-\sum_{<p,pm,n>} \mu_3(p)\mu_3(p')-\lambda \sum\limits_{p} \mu_{1}(p) \equiv H_E + \lambda H_B, \nonumber\\
& ~~~~~~=\lambda\bigg[ -\sum\limits_{p}\mu_{1}(p)-\frac{1}{\lambda}\sum\limits_{<p,p'>} \mu_3(p)\mu_3(p')\bigg]
 \label{dhz}
\end{align}
 In (\ref{dhz}) $\sum_{<p,p'>}$ denotes the sum over the nearest neighbor plaquettes. 
 Note that the original fundamental non-interacting electric field terms are now described by nearest neighbor  interacting electric scalar potentials. 
 The non-interacting magnetic fields, on the other hand, have now acquired the fundamental  status. 
Thus the two gauge-spin descriptions: 
 $$\{\sigma_1(l); \sigma_3(l)\} \leftrightarrow  
 \{\mu_{1}(p); \mu_3(p)\}$$ are related by duality. Further,  $Z_2$ lattice gauge theory at coupling $\lambda$ is mapped into $Z_2$ spin model 
 at coupling  $(1/\lambda)$, i.e, $$H^{{}^{Z_2}}_{gauge}(\lambda)  \simeq \lambda ~H^{{}^{Z_2}}_{spin}({1}/{\lambda}).$$
We have used $\simeq$ above to emphasizes that this equivalence is only within the physical Hilbert space ${\cal H}^p$.    

\subsubsection{\bf $Z_2$  Magnetic disorder operator} 
\label{sz2od} 
The dual spin model (\ref{dhz}) on an infinite lattice has global $Z_2$ invariance:
   \begin{align}
   \mu_{1}(p) 
   \rightarrow 
   \mu_{1}(p), ~~~~
   \mu_3(p)
  \rightarrow 
   -\mu_3(p), ~~\forall p \in \Lambda.
   \label{z2gs}
   \end{align}
  Its  generator 
   ${G}_\Lambda \equiv \prod\limits_{p\in \Lambda} \mu_{1}(p)$ leaves the Hamiltonian (\ref{dhz}) invariant:  ${ G}_\Lambda  H { G}^{-1}_\Lambda =H$.  
  Unlike the initial $Z_2$ gauge symmetry of $Z_2$ gauge theory, the  global $Z_2$ symmetry of the dual spin model (\ref{dhz}) is the symmetry of the spectrum. Being independent of  gauge invariance, it allows the Ising spin model (\ref{dhz}) to be magnetized through spontaneous symmetry breaking for $\lambda << 1$.  As a consequence of duality:
  \begin{align} 
\hspace{-0.2cm} \Big<\mu_{1}(p)\Big>_{{}_{H^{z_2}_{spin}({1}/{\lambda})}}   = 
        \Big<&\sigma_3(l_1)\sigma_3(l_2)\sigma_3(l_3)\sigma_3(l_4)\Big>_{{}_{H^{z_2}_{gauge}(\lambda)}}\nonumber \\
 \Big<\mu_3(m,n)\Big>_{H^{z_2}_{spin}(1/\lambda)}  & =
   \left<\prod_{n'=n}^{N}\sigma_1(m,n')\right>_{H^{z_2}_{gauge}(\lambda)}
   \label{dcf} 
   \end{align}
   \begin{figure}[t]
    	\centering
    	\includegraphics[scale=.7]{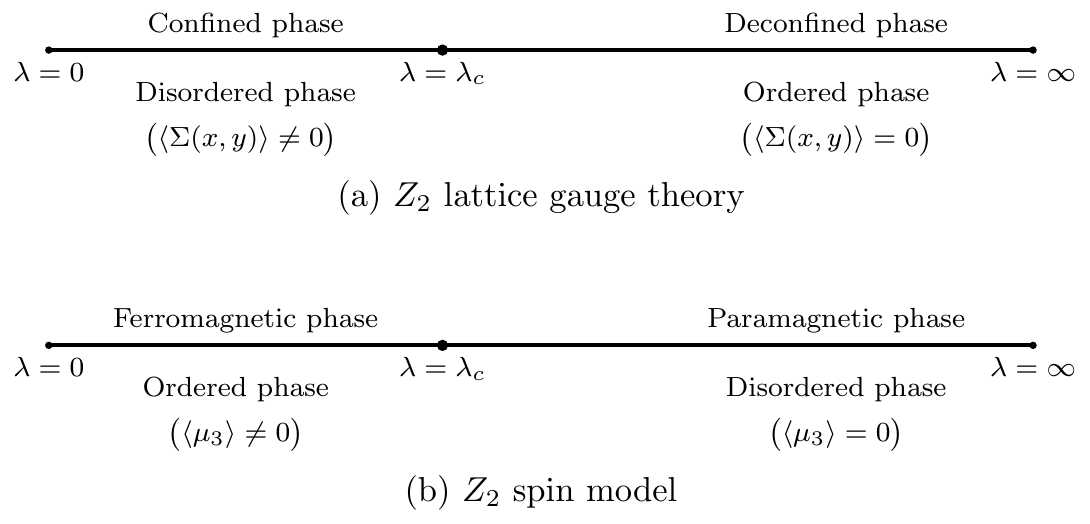}
    	\caption{Duality and order, disorder in (a) $(2+1)$ dimensional $Z_2$ lattice gauge theory, (b) $(2+1)$ dimensional Ising model. The confining ($\lambda << 1$) and deconfining ($\lambda >>1$) phases of $Z_2$ lattice gauge theory correspond to the ferromagnetic and paramagnetic phases of the Ising spin model \cite{kogrev,fradsuss,horn,savit}.} 
    	\label{z2disorder} 
    \end{figure}
   
The above two equations describe the relationship between order and disorder in the gauge and the dual spin system. Note that  we always measure  order or disorder  with respect to the potentials.   
The first relation above states that at low temperature or large coupling   $\lambda >> 1$,   
the gauge system is in ordered phase. This is because   all  magnetic vector potentials $(\sigma_3(l))$ are aligned (close to unity) leading to  $\sigma_3(l_1)\sigma_3(l_2)\sigma_3(l_3)\sigma_3(l_4) 
\approx 1$. This is the free phase of $Z_2$ gauge theory mentioned in the introduction with Wilson loop following  perimeter law:
\begin{align} 
&\langle W_{[{\cal C}]} \rangle \equiv \langle \prod_{l \in {\cal C}} \sigma_3(l) \rangle  
= exp \Big(-{\lambda}^{-2}~{\textrm Perimeter}{({\cal C})}\Big),\nonumber \\
&\hspace{2cm} \lambda >> 1.
\label{parimeter}
\end{align} 
 However, the dual spin system is now at high temperature. It is in the disordered phase as the dual electric scalar potential or the spin values $\mu_3(p)=\pm 1$  are equally  probable. 
On the other hand, at small coupling ($\lambda <<1$), the spin system is ordered with all electric scalar potentials  aligned to the value $\mu_3(p)= +1$ or $-1$. The  gauge system is now disordered as the two values of the  magnetic vector potentials $\sigma_3(l) = \pm1$ are equally probable. This  is the confining phase with the $Z_2$ Wilson loop  $W_{[{\cal C}]}$ around a closed curve ${\cal C}$ following the  area law:
 \begin{align} 
 &\langle W_{[{\cal C}]} \rangle \equiv \langle \prod_{l \in {\cal C}} \sigma_3(l) \rangle  
 \sim \left({\lambda}\right)^{Area({\cal C})} = exp \Big(-|ln \lambda|~{\textrm Area}{({\cal C})}\Big),\nonumber \\
 &\hspace{2cm}\lambda << 1.
 \label{areal}
 \end{align} 
 \begin{figure*}[t]
  \includegraphics[scale=.8]{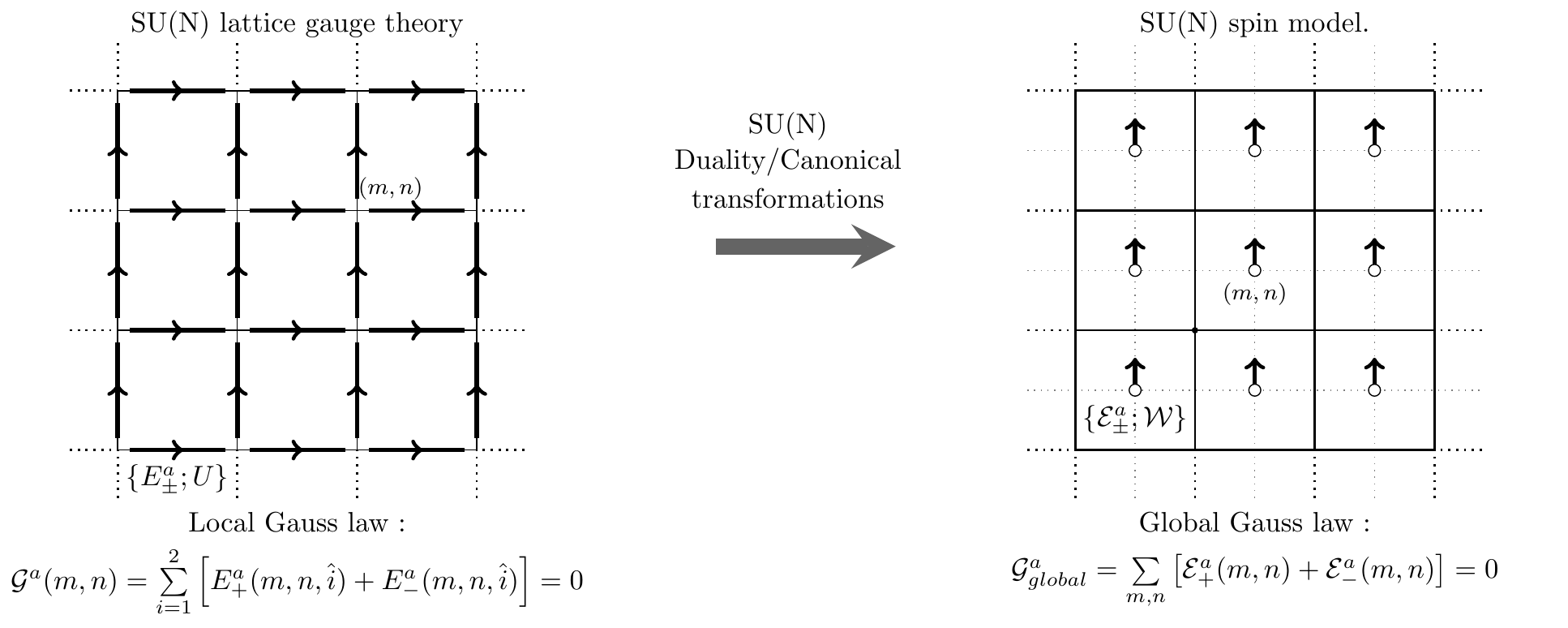}
  \caption{Duality between SU(N) lattice gauge theory and an SU(N) spin model. Unlike the corresponding $Z_2$ duality in Figure \ref{nglc}-a,b,  global SU(N) Gauss law constraints
  at the origin remain unsolved due to the non-abelian
  nature of the gauge transformations.}
  \label{sundf} 
  \end{figure*}
The disorder in the gauge system is  the order in the dual spin system which is measured by  the expectation value of electric scalar potential $\mu_3(p)$. It is  a (non-local) product 
of   the original link electric fields which flip the 
magnetic vector potentials $\sigma_3(l)$ along an infinite path. This is shown  in Figure \ref{rellz2}-a.  For latter convenience and comparisons with SU(N) results, the $Z_2$ disorder operator is relabeled as: 
\begin{align} 
{\Sigma}(m,n) \equiv \mu_3(m,n)  = \prod_{n'=n}^{N}~\sigma_1(m,n'). 
\label{z2do} 
\end{align}
 Just like in the case of Kramers-Wannier duality \cite[]{km,fradsuss,kogrev,horn}, the disorder operator $\Sigma(m,n)$ in $Z_2$ gauge theory acting on an ordered state creates a kink state \cite[]{fradsuss,kogrev} which is orthogonal to the original ordered state. Note that a kink at plaquette 
 $p$  is a  magnetic vortex at $p$  in the original gauge language.  Therefore the expectation value of the disorder operator in an ordered state (no kinks or vortices) is $0$. Below the critical point $\lambda_c$, its expectation value is non-zero. This is the disordered phase and can be understood in terms of kink or magnetic vortex condensation \cite[]{kogrev}. 
 We therefore obtain: 
 \begin{eqnarray} 
 \left<\Sigma(m,n)\right> 
 \neq  0
 ~~~~~~~~~~~~~~\lambda << 1, 
  \nonumber \\ 
 \left<\Sigma(m,n)\right> 
 = 0  ~~~~~~~~~~~~~~\lambda >> 1.
 \label{z2ndo} 
 \end{eqnarray}
 The $Z_2$ gauge-spin duality and the phase 
 diagrams are shown in Figure \ref{z2disorder}. 
 Note that  the disorder operator $\Sigma(m,n)$ 
 is gauge invariant as it  commutes with the local  
 Gauss law operators ${\cal G} (n)$.
We further define $\mu_3(m,n) \equiv e^{i \pi {\cal E}(m,n)}$. Using (\ref{z2efmp}), 
we get: 
\begin{eqnarray}
 &&{\Sigma}(m,n)  =  ~~exp{~i\Big(\tiny{\pi~ {\cal E}(m,n)}\Big)} =\nonumber  \\
& & exp ~i  \Big(\pi \footnotesize{\sum_{n'=n}^{N} E(m,n'; \hat 1)}\Big) \equiv {\Sigma}_{\pi}(m,n).  
\label{z2do2} 
\end{eqnarray}  
The order-disorder algebra is obtained by using  the anti-commutation relation between $\sigma_1(l)$ and  $\sigma_3(l)$:
\begin{align}
 W_{[{\cal C}]} ~\Sigma_\pi(m,n)= \left(-1 \right)^q  
 ~\Sigma_\pi(m,n) ~ W_{[{\cal C}]}.
\label{wtha} 
\end{align} 
As  ${\cal C}$ is a closed loop: 
$q=1$ if the point $(m,n)$ is inside ${\cal C}$ and $q=0$ if $(m,n)$ is outside ${\cal C}$. This can be generalized to more complicated curves  where $q$ equals the winding number which is the number of times the curve ${\cal C}$ winds around the plaquette at $(m,n)$. The algebra (\ref{wtha}) is the standard Wilson-'t Hooft loop algebra for the simplest $Z_2$ lattice gauge theory in $(2+1)$ dimensions.
In the next section, we construct SU(N) duality 
transformations and  exploit them to generalize 
(\ref{z2do2}) and  (\ref{wtha}) to SU(N) lattice gauge theory.

\subsection{SU(N) duality and SU(N) Spin Model} 
\label{snalgt}

In this section, we construct SU(N) spin model which is dual to SU(N) lattice gauge theory.  As mentioned earlier, dualities in abelian, non-abelian lattice gauge theories have been  extensively studied in 
the past 
\cite{thmds,hooft,banks,fradsuss,kogrev,savit,horn,sharatram,manu}. 
Most of these studies involve path integral approach
and abelian gauge groups. The duality transformations 
are used  to make the  compactness of the abelian and non-abelian  gauge groups manifest in the form of 
topological (magnetic monopoles) degrees of freedom 
\cite{banks}.  Our purpose in this section  is to show  
that SU(N) lattice gauge theory can be constructively dualized 
like $Z_2$ lattice gauge theory in the previous section.  This is illustrated in Figure \ref{sundf}.
The SU(N) dual (spin) operators  also lead to a new SU(N) disorder operator discussed  
in section \ref{sdo}.
We directly motivate the SU(N) results through the $Z_2$ lattice gauge theory duality discussed in the previous section.
All algebraic details of  SU(N) canonical transformations can be  found  in \cite{msprd}. 
For the sake of comparison and convenience, all  initial and  final dual spin operators involved in $Z_2$ and $SU(N)$ lattice gauge theories are 
shown in Table-1. The abelian $U(1)$ results 
can be easily obtained by ignoring all non-abelian terms from the duality transformations at the end.

The basic operators involved in the  Kogut-Susskind Hamiltonian formulation of SU(N) lattice gauge theories 
are $SU(N)$ flux operators $U(\vec{n};\hat i)$  and the corresponding left, right electric fields $E^a_{+}(\vec{n}; \hat i)$ and $E^a_-(\vec{n}+\hat i; \hat i)$ and  on every link 
\cite{ks}. They  satisfy the following canonical commutation relations: 
\begin{subequations}
\begin{align}
\left[E^{a}_{+}(\vec n;\hat i),U_{\alpha\beta}(\vec n; \hat i)\right]  =   - \left(\frac{\lambda^a}{2} U(\vec n; \hat i)\right)_{\alpha\beta}, 
\nonumber\\ 
\left[E^{a}_{-}(\vec n+\hat i; \hat i),U_{\alpha\beta}(\vec n; \hat i)\right] =  \left(U(\vec n; \hat i)\frac{\lambda^a}{2}\right)_{\alpha\beta}  
\label{ccr11}
\end{align} 
\begin{align}
\left[E^{a}_{+}(\vec n;\hat i),E^b_{+}(\vec n; \hat i)\right]  = i f^{abc} E^{c}_{+}(\vec n;\hat i),  \nonumber \\    
\left[E^{a}_{-}(\vec n;\hat i),E^b_{-}(\vec n; \hat i)\right]  = i f^{abc} E^{c}_{-}(\vec n;\hat i).
\label{ccrl2}
\end{align} 
\end{subequations}
In (\ref{ccr11}) and (\ref{ccrl2}), $\lambda^{a} (a=1,2, \cdots , N^2-1)$ are the representation  matrices in 
the fundamental representation of SU(N) satisfying $Tr \left(\lambda^a \lambda ^b\right) = \frac{1}{2}\delta^{ab}$ 
and $f^{abc}$ are the SU(N) structure constants. 
$E_-^a(\vec n+\hat i; \hat i)$ and $ E_+^a(\vec n; \hat i)$ are the generators of right and left gauge transformations on the link flux operator $U(\vec n; \hat i)$.
The left and the right electric fields are not independent and are related by:
\begin{align}  
E_-^a(\vec n+\hat i; \hat i) = - R_{ab}(U^\dagger(\vec n; \hat i)) E_+^b(\vec n; \hat i); \nonumber\\~~~~R_{ab}(U(\vec n; \hat i)) 
\equiv \frac{1}{2} Tr \left(\lambda^a U (\vec n; \hat i) \lambda^b U^{\dagger} (\vec n; \hat i)\right). 
\label{lrefr}
\end{align} 
The rotation operator R satisfies $R^{T} R = R R^T =1$.  The local SU(N) gauge transformations rotate the link operators and the electric fields as: 
\begin{align}
E_{\pm}(\vec n; \hat i) \rightarrow \Lambda(\vec n) ~E_{\pm}(\vec n; \hat i)~ \Lambda^{\dagger}(\vec n),\nonumber\\
U(\vec n; \hat i) \rightarrow \Lambda(\vec n)~U(\vec n; \hat i)~\Lambda^{\dagger}(\vec n+\hat i)
\label{gt1n}
\end{align}
the generators of SU(2) gauge transformations at any lattice site n are: 
\begin{eqnarray} 
{\cal G}^{a}(\vec n) = \sum_{i=1}^{d=2}\Big(E_{-}^{a}(\vec n; \hat i) + E_{+}^{a}(\vec n; \hat i)\Big), ~~~~~~~\forall ~~\vec n, ~a. 
\label{su2gln} 
\end{eqnarray}
Therefore there is a Gauss law constraint ${\cal G}^{a}(\vec n)|\psi\rangle_{phys} =0$ at each lattice site $\vec n$, where $|\psi\rangle_{phys}$ is any physical state. 
The Hamiltonian is
\begin{align}
H&=\frac{g^2}{2}\sum\limits_l E^a(l)E^a(l) +\frac{1}{2g^2} \sum\limits_{p}\Big(2 N-\left(Tr~ U_p+h.c\right)\Big) \nonumber \\
&~~~~~~~~~~~~~~~\equiv ~\left(g^2 H_E+\frac{1}{g^2}H_B\right).
\label{hamks}
\end{align}
Above, $l$ and $p$ refer to links and plaquettes on the lattice. $U_p$ is the product of link operators corresponding to the links along a plaquette.  $g^2$ is the coupling constant. Like in $Z_2$ gauge theory Hamiltonian (\ref{isingH}), all interactions are contained in the magnetic part $H_B$  of the Hamiltonian. The electric part $H_E$, with 
no interactions, can be easily diagonalized 
leading to gauge invariant strong coupling 
expansion in terms of loop states \cite{ks,manu}.  
After duality in the next section, like $Z_2$ 
lattice gauge theory in section \ref{z2dd}, 
their  roles  
will 
be reversed.

 
\begin{figure}[t]
   \centering
    \includegraphics[scale=0.7]{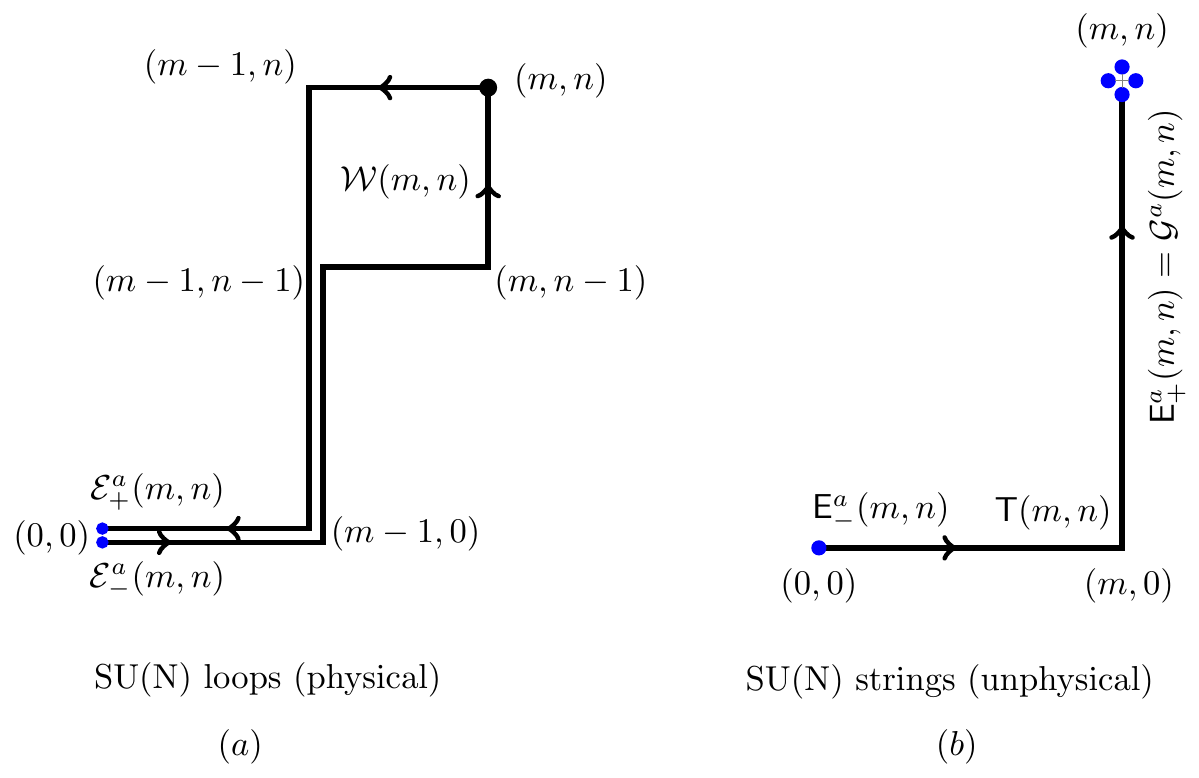}
    \caption{The physical SU(N) spin conjugate pairs 
    $\{{\cal E}_{\pm}; {\cal W}_{\alpha\beta}\}$ 
    and the unphysical SU(N) string conjugate pairs $\{{\sf E}_\pm(n); {\sf T}(n)\}$ dual to SU(N) lattice gauge theory are shown in (a) and (b) respectively. Like in $Z_2$ case in Figure \ref{dual}, we label the SU(N) spin operators by their top right corners and the SU(N) string operators  by their endpoints.  The strings decouple from the physical Hilbert space as ${\sf E}_+^a(m,n) \approx 0$ by the Gauss law constraints 
     in ${\cal H}^p$. 
    } 
    \label{dualsun}
   \end{figure}
      
      \subsubsection{\bf Physical sector and SU(N)  dual  potentials}
      \begin{footnotesize}
  \begin{table*}
              \label{ttt1} 
              \begin{center}
              \begin{tabular}{|c|c|c|c|}
              \hline
              \multicolumn{2}{|c|}{\textbf{$Z_2$ lattice gauge theory}} & \multicolumn{2}{|c|}{\textbf{SU(N) lattice gauge theory}}\\
              \hline
              Gauge Operators & \multicolumn{1}{|c|}{ Dual/Spin Operators} & Gauge Operators & \multicolumn{1}{|c|}{ Dual/Spin Operators} \\ 
              \hline
              & ~~~~  $\left\{ \mu_1(m,n); \mu_{3}(m,n) \right\}$~~~~  &&  ~~~~~  $\left\{{\cal E}_\pm^a(m,n);~{\cal W}_{\alpha\beta}(m,n)\right\}$ ~~~~\\
              &{\small{($Z_2$ Loops/$Z_2$ Ising spins)}} && (SU(N) Loops/SU(N) spins) \\
              & & & \\
              ~$\left\{ \sigma_1(m,n;\hat{i});\sigma_3(m,n;\hat{i}) \right\}$ ~ &  &~~  $\left\{ E_\pm^a(m,n;\hat{i}); U_{\alpha\beta}(m,n;\hat{i}) \right\}$ ~~  & ~~~~   \\ &&& \\~~~~ &~~~~ 
               $\left\{ \bar{\sigma}_1(m,n); \bar{\sigma}_3(m,n) \right\}$ ~~~~ &~~~~  &~~~~ $\left\{{\sf E}_\pm^a(m,n);~ {\sf T}_{\alpha\beta}(m,n)\right\}$ \\
               &~~~~ (Frozen $Z_2$ Strings)~~~~ &&~~~~ (Frozen $SU(N)$  Strings)\\
              \hline
              \end{tabular}
              \end{center}
              \caption{The basic  conjugate operators of the original and the  dual $Z_2$,  SU(N) gauge theories 
               in $(2+1)$ dimensions.}
              \end{table*}
              \end{footnotesize} 
              
              We now define the dual SU(N) spin and SU(N) string operators analogous to the $Z_2$ spins and strings in (\ref{plaqop1}), (\ref{plaqop2}) and  (\ref{z2duality1}), (\ref{z2duality2})  respectively.  They 
              are pictorially
              described in Figure \ref{dualsun}-a,b respectively.  
    Due to the non-abelian nature of the electric field and the flux operators, the SU(N) duality relations have additional non-abelian structures \cite{msprd}.  To begin with, the  ${\cal N}$ SU(N)  Gauss law constraints at ${\cal N}$ different lattice sites are all mutually independent.  In other words, identities like (\ref{z2gli}) do not exist. As a result, there is a global SU(N) invariance in the SU(N) spin model corresponding to the gauge transformations at the origin. All dual operators transform covariantly under this global SU(N). As shown in Figure \ref{dualsun}, the SU(N) duality transformations 
    involve parallel
   transports  from the origin to the site of the dual operators.  The string flux operator ${\sf T}(m,n)$  at a lattice site $(m,n)$ (analogous to $\bar \sigma_3(m,n)$ in the $Z_2$ case)  is defined  
           through the path $(0,0) \rightarrow (m,0) \rightarrow (m,n)$:
              \begin{figure}[b]
              	\includegraphics{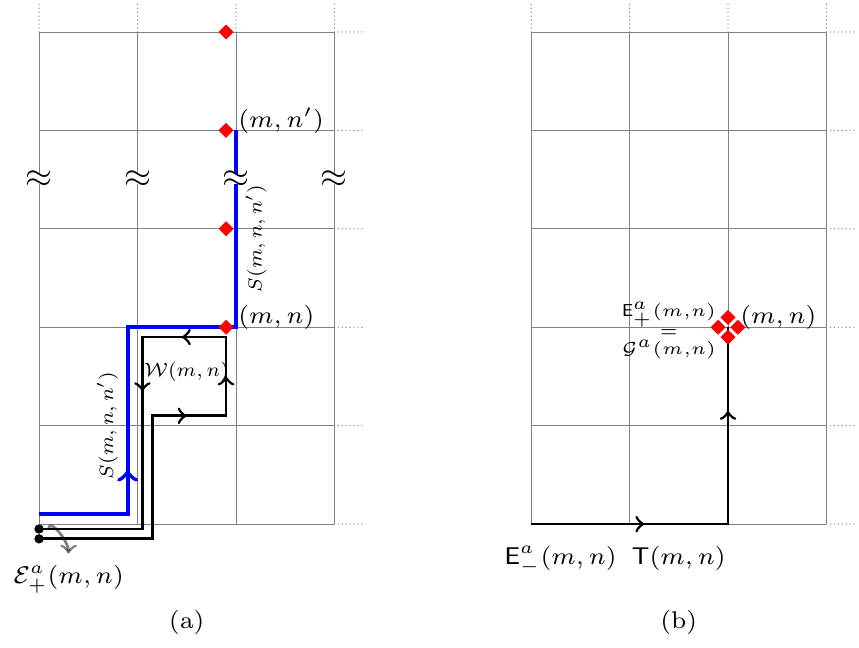}
              	\caption{The non-local relations in  SU(N) duality transformations 
              		and the Gauss law constraints.
              		(a) We show the relations (\ref{sunplaqe}) expressing  ${\cal E}_+^a(m,n)$ as the sum of $E_-^b(m,n'; \hat 1)$. 
              		The Kogut-Susskind electric fields and 
              		and the plaquette loop electric fields
              		are denoted by {\tiny{\color{red}$\blacklozenge$}} and $\bullet$ respectively.
              		In 
              		(b) we show the SU(N) Gauss law constraints (\ref{splo}). 
              		The corresponding $Z_2$ illustrations are  in Figure 
              		\ref{rellz2}-a,b,c.}
              	\label{frelsun}
              \end{figure}
           \begin{subequations}
             \begin{align} 
        {\sf T}(m,n)  = \left(\prod_{m^\prime=0}^{m}U(m^\prime,0; \hat{1})
                \prod_{n^\prime=0}^{n} U(m,n^\prime; \hat{2})\right), 
                \label{sunstring}
                \end{align}
                \begin{align}
              \hspace{-0.2cm}   {\sf E}^a_+(m,n) = {\cal G}^a(m,n) \approx 0, 
                 \label{sungl}                 
                \end{align}
                \end{subequations}
                These strings and their electric fields 
                ${\sf E}^a_+(m,n)$ are shown in Figure \ref{dualsun}-b and Figure \ref{frelsun}-b
                respectively.  The relations (\ref{sunstring}) and (\ref{sungl})
                are the SU(N) analogues  of the $Z_2$ string relations (\ref{z2duality1}) and 
                (\ref{z2duality2}) respectively.
     The  dual SU(N) spin and the SU(N)  electric scalar potential operators in terms of the original Kogut-Susskind operators are defined  \cite{msprd} as
\begin{subequations}
     \begin{align} 
     \hspace{-0.2cm} {\mathcal W}(m,n)& ={\sf T}(m\hspace{-0.1cm}-\hspace{-0.1cm}1,n\hspace{-0.1cm}-\hspace{-0.1cm}1)~U_p(m,n)~{\sf T}^\dagger(m\hspace{-0.1cm}-\hspace{-0.1cm}1,n\hspace{-0.1cm}-\hspace{-0.1cm}1), 
     \label{sunplaq}
     \end{align}
     \begin{align}
   \hspace{-0.3cm}  {\mathcal E}_+^a(m,n)& = \sum\limits_{n'=n}^{\sf N} R_{ab}(S(m,n;n^\prime))E_-^b(m,n'; \hat{1}). 
     \label{sunplaqe}
     \end{align} 
     \end{subequations} 
     The two  operators  in (\ref{sunplaq}) and  (\ref{sunplaqe}) are the non-abelian  
     extensions  of the two $Z_2$ dual operators  defined  in (\ref{plaqop1}) and (\ref{plaqop2}) 
     respectively.
In (\ref{sunplaq}), (\ref{sunplaqe}),  the plaquette operator $U_p(m,n)$ and the parallel transport $S(m,n;n')$ are defined as
     \begin{align}
     \label{dew}
     &U_p(m,n)= U(m-1,n-1;\hat{1})~U(m,n-1;\hat{2}) \nonumber\\ 
   &~~~~~~~~U^\dagger(m-1,n;\hat{1})~U^\dagger(m-1,n-1;\hat{2}),  \\ 
   & S(m,n;n^\prime)  \equiv {\sf T}(m\hspace{-0.1cm}-\hspace{-0.1cm}1,n)~U(m\hspace{-0.1cm}-\hspace{-0.1cm}1,n;
   \hat{1})~\prod_{q=n}^{n'}~U(m,q; \hat{2}). \nonumber 
   \end{align} 
    The relation (\ref{sunplaq}) defines the SU(N) magnetic field operator as a fundamental operator. 
   The second relation (\ref{sunplaqe}) defines SU(N)  electric scalar potential ${\cal E}^a(m,n)$ which is dual to the original magnetic vector potential.
    The appearance of 
      the ${\sf T}(m,n)$ and $S(m,n;n')$ in (\ref{sunplaq}) ad (\ref{sunplaqe})
      is due to the non-abelian nature of the operators. These parallel transports from the origin  are required to have consistent gauge  transformation properties of the SU(N) magnetic 
      fields and the  SU(N) electric scalar potentials (see (\ref{wgtao})).   
      
   The dual or loop operators satisfy the expected non-abelian  duality or quantization rules:
   \begin{subequations}
     \begin{align}
   \left[{\cal E}^{a}_{-}(m,n),{\cal W}_{\alpha\beta}(m,n)\right]  &=   - \left(\frac{\lambda^a}{2} {\cal W}(m,n)\right)_{\alpha\beta}, \nonumber\\  
    \left[{\cal E}^{a}_{+}(m,n),{\cal W}_{\alpha\beta}(m,n)\right] &=  \left({\cal W}(m,n)\frac{\lambda^a}{2}\right)_{\alpha\beta},
     \label{ccrls1} 
     \end{align} 
     \begin{align}  
  \left[{\cal E}^{a}_{-}(m,n),{\cal E}_{-}^b(m,n)\right] & =   if^{abc} {\cal E}^{c}_{-}(m,n),\nonumber\\
   \left[{\cal E}^{a}_{+}(m,n),{\cal E}_{+}^b(m,n)\right] & =   if^{abc} {\cal E}^{c}_{+}(m,n).
   \label{ccrls2}   
   \end{align}
   \end{subequations}  
  Further, the two electric fields are related through parallel transport and commute:
   \begin{eqnarray}
   & {\cal E}_-^a(m,n) \equiv - R_{ab}({\cal W}^\dagger(m,n)) {\cal E}_+^b(m,n) \nonumber\\ 
   & =>
    \left[{\cal E}^a_-(m,n),{\cal E}^b_+(m,n) \right] =0. 
    \label{tefr}
    \end{eqnarray}  
   The quantization relations (\ref{ccrls1}), (\ref{ccrls2}) and (\ref{tefr}) are exactly similar to the 
   original quantization  rules (\ref{ccr11}) and 
   (\ref{ccrl2}) respectively.  Thus
   the electric field operator $E^a(m,n; \hat i)$ and  the magnetic vector potential operator $U_{\alpha\beta}(m,n;i)$ have been replaced by their dual 
    electric scalar potential
   ${\cal E}^a(m,n)$ and  the dual magnetic field operator ${\cal W}_{\alpha\beta}(m,n)$.  
   This is similar to $Z_2$ lattice gauge theory duality where $\{\sigma_1(m,n); \sigma_3(m,n)\}$ 
   get replaced by $\{\mu_3(m,n); \mu_1(m,n)\}$. 
   We again emphasize that    ${\cal E}^a(m,n)$ defines 
   the dual electric scalar potential as it  is conjugate to the fundamental magnetic flux operator ${\cal W}_{\alpha\beta}(m,n)$.  
 
\subsubsection{\bf Unphysical sector and SU(N) string  operators}      

The unphysical sector, representing the gauge degrees 
of freedom,  consists of the string flux operators ${\sf T}(m,n)$ in (\ref{sunstring}) and their conjugate electric fields ${\sf E}^a(m,n)$ in (\ref{sungl}). They satisfy the canonical quantization relations:
\begin{subequations}
\begin{align}
\left[{\sf E}^{a}_{-}(m,n),{\sf T}_{\alpha\beta}(m,n)\right]  &=   - \left(\frac{\lambda^a}{2} {\sf T}(m,n)\right)_{\alpha\beta},\nonumber \\  
 \left[{\sf E}^{a}_{+}(m,n),{\sf T}_{\alpha\beta}(m,n)\right] &=  \left({\sf T}(m,n)\frac{\lambda^a}{2}\right)_{\alpha\beta}
 \label{ccrss1} 
 \end{align}
 \begin{align} 
\left[{\sf E}^{a}_{-}(m,n),{\sf E}_{-}^b(m,n)\right]  &=   if^{abc} {\sf E}^{c}_{-}(m,n), \nonumber\\
\left[{\sf E}^{a}_{+}(m,n),{\sf E}_{+}^b(m,n)\right]  &=   if^{abc} {\sf E}^{c}_{+}(m,n).
\label{ccrss2}   
\end{align}
\end{subequations} 
Again, the operators ${\sf T}_+^a$ and ${\sf T}_-^b$ are related through parallel transport and commute amongst themselves:
 \begin{eqnarray}
   & ~~{\sf E}_-^a(m,n) \equiv - R_{ab}({\sf T}^\dagger(m,n)) {\sf E}_+^b(m,n) \nonumber\\ 
   & =>~~
    \left[{\sf E}^a_-(m,n),{\sf E}^b_+(m,n) \right] =0. 
    \label{tefre}
    \end{eqnarray}
    The right string electric fields are 
\begin{eqnarray}
{\sf E}^a_+(m,n) &= \sum_{i=1}^{2} \left[ E^a_-(m,n;\hat i)  + E^a_+(m,n;\hat i)\right]\nonumber\\
& = {\cal G}^a(m,n) \approx 0, ~~\forall (m,n) \neq (0,0).
\label{splo}  
\end{eqnarray}
Thus as in $Z_2$ lattice gauge theory,  the SU(N) Gauss law constraints freeze all SU(N)  string degrees  of freedom. This is shown in Figure \ref{dualsun}-b. 
As a consequence, all strings (or gauge degrees of freedom) completely decouple from the theory. 

\subsubsection{\bf The residual Gauss law} 
\label{rgl}
Unlike $Z_2$ lattice gauge theory, the SU(N) Gauss law at the origin is independent of the SU(N) Gauss laws at other sites.
In other words,  the abelian  identity 
(\ref{z2gli}) has no non-abelian  analogue. 
Under this residual global gauge invariance at the origin $\Lambda \equiv \Lambda(0,0)$,  all loop operators transform like adjoint matter fields:
 \begin{align}
 {\cal E}_{\pm}(p)  \rightarrow  \Lambda ~{\cal E}_{\pm}(p) ~\Lambda^\dagger,~~~
  {\cal W}(p)  \rightarrow \Lambda ~{\cal W}(p) ~ \Lambda^\dagger.
  \label{wgtao}
 \end{align}
 Above, $ {\cal E}_{\pm}(p) \equiv {\cal E}_{\pm}(m,n),  {\cal W}(p) \equiv {\cal W}(m,n)$ and $\Lambda \equiv \Lambda(0,0)$ is the gauge 
 transformation at the origin.
 This global invariance at the origin is fixed  by the $(N^2-1)$ global SU(N) Gauss laws:
 \begin{align}
 &~{\cal G}^a \equiv {\cal G}^a(0,0)= E^a_+(0,0;\hat{1})+E^a_-(0,0;\hat{2}) \nonumber \\
 &=\sum_{m=1}^{\sf N} \sum_{n=1}^{\sf N} \left[\underbrace{{\sf E}^a_-(m,n)}_{=0}+\underbrace{{\cal E}^a_+(m,n)+{\cal E}^a_-(m,n)}_{\equiv {\mathbb L}^a(m,n)}\right]\nonumber \\  
& ~~~~~~~~\equiv \sum_{m=1}^{\sf N}\sum_{n=1}^{\sf N} {\mathbb L}^a(m,n) =0.
\label{resgl} 
 \end{align}
 In (\ref{resgl}), the total left and right electric flux operators on a plaquette located at $p=(m,n)$ are donated by ${\mathbb L}^a(m,n)$ and  equations (\ref{tefre}), (\ref{splo}) are used  to get ${\sf E}^a_-(m,n) =0$. In Appendix \ref{nagls} we 
 show that the dual SU(N) electric scalar potentials,
 like $Z_2$ electric potentials in (\ref{z2glprfe}), 
 solve the SU(N) 
 Gauss law constraints away from the origin 
 and lead to (\ref{resgl}) at the origin.  
 The residual global constraints (\ref{resgl}) can be 
 solved by using the angular momentum or spin network basis \cite{sharatram,manu,msplb,msprd}. Note that 
 in the abelian U(1) case there is no residual Gauss law as ${\mathbb L}^a(m,n) \rightarrow {\mathbb L}(m,n)
\equiv {\cal E}_+(m,n) + {\cal E}_-(m,n) \equiv 0$.
 
\subsubsection{\bf Inverse relations}
\noindent The inverse flux operator relations, analogous to the $Z_2$ relations (\ref{iroel}), 
are 
\begin{eqnarray}
U(m,n; \hat{1})   =  {\sf T}^\dagger(m,n)~{\cal W}(m+1,n)~{\cal W}(m+1,n-1)\cdot \nonumber \\ 
\cdots {\cal W}(m+1,1)~{\sf T}(n+1,y), ~~~~~~~~~\nonumber  \\
U(m,n; \hat{2}) = ~T(m,n+1)~T^\dagger(m,n). \hspace{2cm} \label{uf}
\end{eqnarray}
The inverse electric field relations, analogous to the $Z_2$ electric field relations (\ref{iroel2}), are
\begin{eqnarray}
E_+^a(m,n; \hat{1}) =  R_{ab}({\sf T}(m,n))\bigg\{{\cal E}_-^b(m+1,n+1) ~~~~~~~\nonumber \\
+ ~{\cal E}_+^b(m+1,n)+\delta_{n,0}\sum_{\bar m=m+2}^L\sum_{\bar n=1}^{\sf N}{\mathbb L}^b(\bar m,\bar n)\bigg\},  ~~\nonumber \\
E_+^a(m,n; \hat{2}) = 
R_{ab}({\sf T}(m,n))\bigg\{{\cal E}_+^b(m+1,n+1) +\nonumber ~~~~\\
R_{bc}({W}(m,n)){\cal E}_-^c(m,n+1)
+\sum\limits_{\bar n=n+2}^{\sf N} {\mathbb L}^b(m+1,\bar n)\bigg\}. 
\label{kstow}
\end{eqnarray}
In the last step in (\ref{kstow}) we have defined, 
$R_{ab}({W}(m,n)) 
\equiv R_{ab}\big(
{\cal W}(m,n){\cal W}(m,n-1)\cdots {\cal W}(m,1)
\big), {\mathbb L}^a(m,n) \equiv \left({\cal E}_-^a(m,n) +{\cal E}_+^a(m,n)\right)$.
 In the abelian U(1) case (\ref{kstow}) involves 
 only the nearest neighbor loop electric fields as there are no color indices  ${\mathbb L}^a \rightarrow {\mathbb L} \equiv 0$ and $R_{ab}({U})\rightarrow 1$. 

\subsubsection{\bf SU(N) dual dynamics}
\label{ssundd} 
\begin{figure*}
       \includegraphics[scale=1.0]{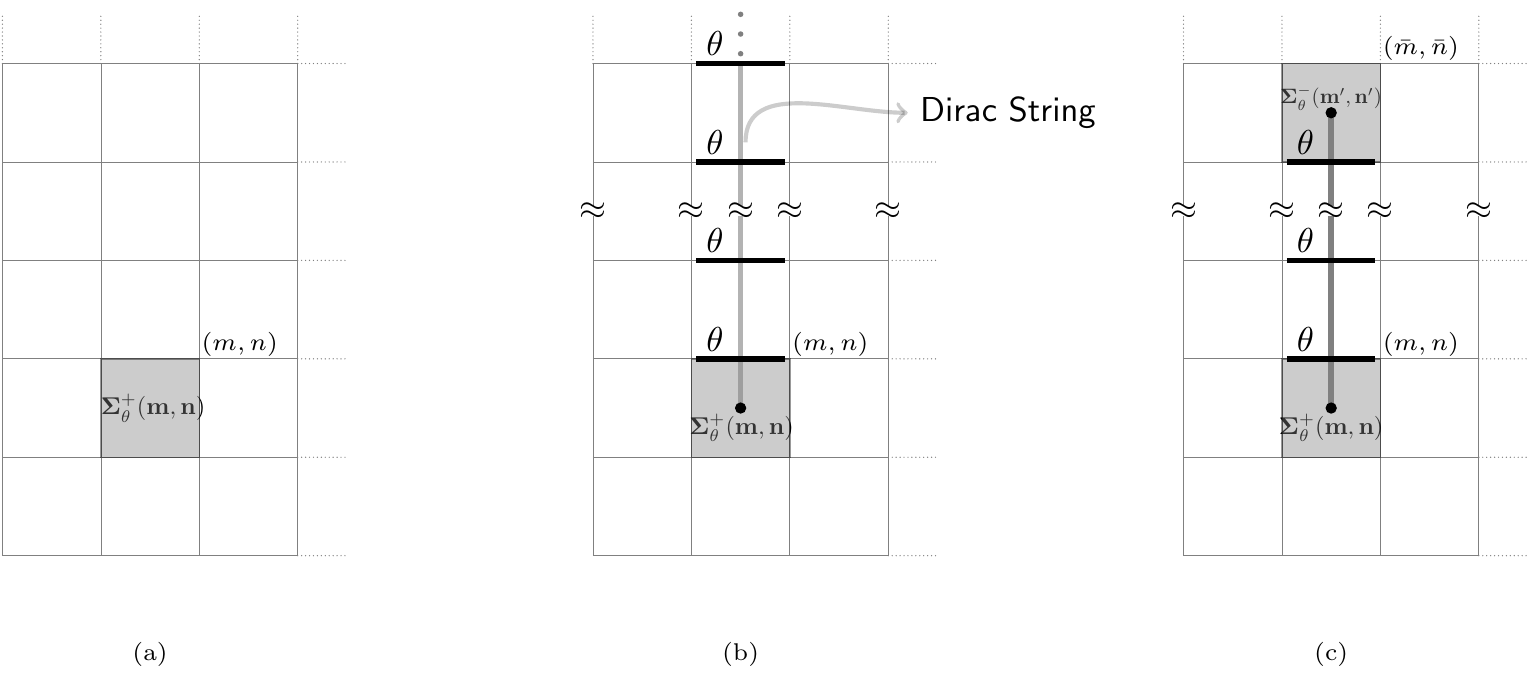}
      \caption{The 
      disorder operator $\Sigma_\theta^+(m,n)$  creating a magnetic  vortex at $(m,n)$ 
       (a) in the dual spin description, (b) in the original Kogut-Susskind gauge description but now with an infinitely long Dirac string, (c) a vortex-anti-vortex pair connected through a finite  length Dirac string.
      The dark heavy horizontal links  across the Dirac strings in (b) and (c) represent rotations of the Kogut-Susskind link operators $U(m-1, n';\hat 1), n' \ge n$ by $\theta$. More details are given in Appendix \ref{ids}.} 
      \label{mo} 
      \end{figure*} 
The Hamiltonian of pure SU(N) gauge theory in terms of the dual operators is 
{\footnotesize 
\begin{align}
& H  =  \sum_{m,n \in \Lambda} \frac{g^2}{2}\Bigg\{\Big[\vec{\cal E}_-(m+1,n+1)+\vec{\cal E}_+(m+1,n) + {\Delta}_{XY}(m,n)\Big]^2  
\nonumber \\
&+\Big[\vec{\cal E}_+(m+1,n+1)
+ R_{bc}({W}(m,n))\vec{\cal E}_-^c(m,n+1)  +  \Delta_Y(m,n)\Big]^2\Bigg\}   \nonumber \\ 
& ~~+ \frac{1}{2g^2} \Big(2N - (Tr~ {\cal W}(m,n)+h.c)\Big) \equiv g^2 \tilde H_E + \frac{1}{g^2} \tilde H_B.
\label{loopham} 
\end{align}
}
In (\ref{loopham}) we have defined \cite{msprd},
\begin{footnotesize} 
${\Delta}^a_{XY}(m,n)  \equiv  \delta_{m,0} \sum\limits_{\bar m=m+2}^{\sf N}\sum\limits_{\bar n=1}^{\sf N} {\mathbb L}^a(\bar m, \bar n)$ \end{footnotesize} and   
\begin{footnotesize}   
${\Delta}^a_{Y}(m,n)  \equiv \sum\limits_{\bar n=n+2}^{\sf N} {\mathbb L}^a(m,\bar n)$,
\end{footnotesize}
where ${\mathbb L}(m,n)$ is given in the 
equation (\ref{resgl}).

Thus the SU(N) Kogut-Susskind Hamiltonian in its dual description (unlike the $Z_2$ lattice gauge theory Ising model Hamiltonian)  becomes non-local. The non-localities in (\ref{loopham}) comes from the terms ${\cal R}({\cal W}),{\Delta}^a_{XY}(m,n)$ and ${\Delta}^a_{Y}(m,n)$. But, since ${\cal R}({\cal W}) = 1+ o(g)+o(g^2)+ \cdots$ , ${\mathbb L}$ is  of order $g^n, ~(n \ge 1)$ which implies that $ {\Delta}^a_{XY}(m,n)$ and ${\Delta}^a_{Y}(m,n)$ are both at least of the order of $g$. Therefore we expect that in the $g^2 \rightarrow 0$ continuum limit, these non-local parts can be ignored to the lowest order at low energies. This leads to a simplified local effective Hamiltonian $H_{spin}$ which may describe pure SU(N) gauge theory at low energies, sufficiently well.
    \begin{eqnarray}
    H_{spin}=  \frac{g^2}{2}\left\{\sum\limits_{p=1}^{\cal P} 4\vec{\cal E}^{~2}(p)+\sum_{<p,p'>}\vec{\cal E}_-(p)\cdot \vec{\cal E}_+(p') \right\} + \nonumber  \\
    \frac{1}{2g^2} \Bigg\{ 2N- \left(Tr {\cal W}(p)+h.c\right)\Bigg\} 
    \equiv \frac{g^2}{2} {\tilde H}'_E + \frac{1}{2g^2} {\tilde H}_B. ~~
     \label{weakh}
    \end{eqnarray}
 In (\ref{weakh}), ${<p,p'>}$ is used to show the nearest plaquettes. 
 The above simplified SU(N) spin Hamiltonian $H_{spin}$ describes nearest neighbouring SU(N) spins interacting through their left and right electric fields.
All interactions are now contained in the 'electric part'  $\tilde{H}'_E$ and the magnetic part $\tilde{H}_B \sim Tr{\cal W}(p)$ is a non-interacting term. As a result, the coupling constant of the dual model is the inverse of that of the original Kogut Susskind model: 
\begin{center}
$H^{SU(N)}_{gauge} \big(\frac{1}{g^2}\big) \simeq H^{SU(N)}_{spin}\big({g^2}\big)$.
\end{center}
We have used $\simeq$ above to state that this equivalence is only within the physical Hilbert 
space ${\cal H}^p$. The above relation is SU(N) analogue of the 
$Z_2$ result 
$H^{{}^{Z_2}}_{gauge}(\lambda)  \simeq  H^{{}^{Z_2}}_{spin}(\lambda^{-1})$ discussed earlier. 

Note that the global SU(N) invariance (\ref{wgtao}) 
of the dual SU(N) spin model is to be fixed by imposing 
the Gauss law  (\ref{resgl}) at the origin. 
The  degrees 
of freedom before and after duality  match 
exactly as follows.  We have converted the initial $3{\cal L}$ Kogut-Susskind link operators into $ 3{\cal P}$ plaquette spin operators and $3({\cal N}-1)$
string operators (see Table 1) and ${\cal L} = {\cal P} + ({\cal N}-1)$.  There are ${\cal N}$ 
mutually independent Gauss laws
in SU(N) (but $({\cal N}-1)$ in $Z_2$ case) lattice gauge theory. Out of these, $({\cal N}-1)$  freeze the $({\cal N}-1)$ strings. We are thus left with a single Gauss law constraint (\ref{resgl}) in SU(N) spin model  
after duality and none in $Z_2$ case.      

\subsubsection{\bf  SU(N) Magnetic disorder operator}
\label{sdo} 
 
Exploiting duality transformations, we now construct a SU(N) gauge invariant operator 
which  measure the magnetic  disorder  in 
the  
gauge system \cite{hooft}.  Such disorder operators 
and their correlations in the context of 2-dimensional Ising model and Kramers-Wannier duality have been extensively discussed in the path integral approach by Kadanoff and Ceva  \cite{kadanoff}.  These disorder variables were called magnetic dislocations.

In this section we work with dual SU(2) 
spin model with global SU(2) gauge invariance
(\ref{wgtao}). We construct a SU(2) invariant 
magnetic disorder or vortex operator which creates 
a magnetic vortex on a single plaquette. 
We focus on a single plaquette $p=(m,n)$  in 
$\Lambda$ with fixed
$m$ and $n$ as in Figure (\ref{mo})
and write the SU(2)  
magnetic spin operator ${\cal W}_{\alpha\beta}$ in the magnetic 
field eigen basis as: 
\begin{align}
&\hspace{-0.3cm} {\cal W}{(m,n)} \equiv cos\left(\frac{\omega_{(m,n)}}{2}\right) \sigma_{0} + i \left(\hat {w}_{(m,n)} \cdot \vec \sigma\right) \sin\left(\frac{\omega_{(m,n)}}{2}\right)\nonumber\\ 
& \hspace{1.8cm}\hat w_{(m,n)}\cdot \hat w_{(m,n)} =1.
\label{wmn}  
\end{align}
In (\ref{wmn}), $\omega(m,n)$ are  gauge invariant angles, $\hat w(m,n)$ are the unit vectors  in the group manifold $S^3$ and  $\sigma_0, \vec \sigma$ 
are  the unit,  Pauli matrices. Under global gauge transformation $\Lambda \equiv \Lambda(0,0)$ in (\ref{wgtao}), $(\omega,\hat w)$  transform as: 
\begin{align} 
\hspace{-0.1cm} \omega(m,n)\hspace{-0.1cm} \rightarrow \hspace{-0.1cm}\omega(m,n), ~~\hat w^a(m,n) \rightarrow R_{ab}(\Lambda) ~\hat w^b(m,n). 
\label{gtp}
\end{align} 
Above, $R_{ab}(\Lambda)$ are given in (\ref{lrefr}). We define  two unitary operators: 
\begin{align} 
{\Sigma }^{\pm}_\theta(m,n) \equiv exp ~i \Big({\hat w }(m,n)\cdot {\cal E}_{\pm}(m,n) \phantom{.} {\theta} \Big), 
\label{mfo}
\end{align} 
which are located on a plaquette $p \equiv (m,n)$ as shown in the Figure \ref{mo}-a. They both are  gauge invariant because ${\cal E}_{\pm}^a(m,n)$ and 
 $\hat w(m,n)$ gauge  transform like vectors. 
In other words, 
$\left[{\cal G}^a, \Sigma^{\pm}_\theta(m,n)\right] =0$,
where ${\cal G}^a$ is defined in (\ref{resgl}).
 As the left and right SU(N)  electric scalar potentials  are related through 
(\ref{tefr}),  ${\Sigma}^{\pm}_\theta(m,n)$ are not independent and satisfy: 
\begin{align} 
{\Sigma}^{+}_{\theta}(m,n) ~{\Sigma}^{-}_{\theta}(m,n) =
{\Sigma}^{-}_{\theta}(m,n) ~{\Sigma}^{+}_{\theta}(m,n)
= {\it I}. 
\end{align}  
Above ${\it I}$ denotes the unit operator in the physical Hilbert space ${\cal H}^p$ and $ \Sigma_\theta^-  = \Sigma_{-\theta}^+$.  The physical meaning of the operators $\Sigma_\theta^{\pm}(m,n)$  is simple and exactly similar in spirit as $\Sigma_{\pi}(m,n)$ in (\ref{z2do2}) in the $Z_2$ lattice gauge theory case. The non-abelian electric scalar potentials  ${\cal E}^a_{\pm}(m,n)$ are conjugate to the magnetic flux operators ${\cal W}_{\alpha\beta}(m,n)$. They satisfy the canonical commutation relations (\ref{ccrls1}). Therefore the gauge invariant operator ${\Sigma}^{\pm}_{\theta}(m,n)$ locally and continuously changes the magnetic flux on the plaquette  $p=(m,n)$  as a function of $\theta$ \footnote{This is similar to the role of momentum operator $\hat p$ as a generator of translation in quantum mechanics: $$e^{i~\hat p ~\theta} {\ket x} = \ket{x+\theta}.$$ This should be 
compared with (\ref{mfo}) and (\ref{aabbcc}).}. 
They are 
the magnetic vortex operators. To see this explicitly, we consider   
eigenstates $\ket{\omega(m,n),\hat w(m,n)}$ of 
${\cal W}_{\alpha\beta}(m,n)$ on a single plaquette. These states are explicitly constructed in (\ref{angbasis}) in  Appendix \ref{appb}. 
They satisfy:
\begin{align} 
Tr {\cal W} ~\ket{\omega,\hat w}  = 2\cos\left(\frac{\omega}{2}\right)~ \ket{\omega,\hat w}.
\label{wlev}  
\end{align} 
We have ignored the irrelevant plaquette index $p\equiv (m,n)$ in (\ref{wlev}) as we are dealing with a single plaquette. It is easy to check:
\begin{eqnarray} 
\ket{\omega, \hat w}_{\pm \theta}  \equiv \Sigma_\theta^{\pm} \ket{\omega, \hat w} = \ket{\omega\pm \theta, \hat w},
\label{aabbcc}  
\end{eqnarray}
implying, 
\begin{eqnarray}
Tr~{\cal W} ~|\omega, \hat w \rangle_{\pm \theta} =2~ cos \left(\frac{\omega \pm \theta}{2}\right) ~|\omega, \hat w \rangle_{\pm \theta}.
\label{mfso}
\end{eqnarray}  
The  equations (\ref{wlev}) and (\ref{mfso}) state:
\begin{align} 
\Sigma^\pm_{2\pi} \phantom{}\left(Tr {\cal W}\right)    = -~\left(Tr {\cal W}\right) \phantom{.}\Sigma^\pm_{2\pi}. 
\label{sthr}
\end{align}
We thus recover the standard Wilson-'t Hooft loop 
$Z_2$ algebra \cite{hooft,to} for SU(2) at $\theta =2\pi$. Similarly, we get
\begin{align}  
\left(Tr{\cal W}^q\right) ~
\Sigma^{\pm}_{2\pi} = \left(-1\right)^q~ \Sigma^{\pm}_{2\pi}~\left(Tr{\cal W}^q\right) .
\label{qsthr}
\end{align} 
These relations are analogous to the $Z_2$ results (\ref{wtha}). The operator  $\Sigma_{2\pi}$ is the 
well known SU(2) 't Hooft operator. 
The relations (\ref{sthr}) and (\ref{qsthr}) can 
be easily generalized to  an arbitrary  Wilson loop $W_C$. In the dual spin model any Wilson loop  can  be written in terms of the ${\cal P}$ fundamental loops ${\cal W}_{\alpha\beta}$ as shown in Figure \ref{wlff}:
 \begin{align} 
 W_C = {\cal W}(p_1)~{\cal W}(p_2)~{\cal W}(p_3)\cdots\cdots {\cal W}(p_{n_c}).
 \label{dwl} 
 \end{align} 
 Here $p_1$ is the plaquette operator in the bottom right corner of $C$ and $p_{n_c}$ is the plaquette operator at the left top corner of $C$. Now the  
 factor  $(-1)^q$ in (\ref{qsthr}) is the phase when the curve $C$ winds the magnetic vortex at  $(m,n)$      (created by $\Sigma^\pm_{2\pi}(m,n)$) $q$ number of times. 
 
The plaquette magnetic flux or the vortex operators $\Sigma^{\pm}_{\theta}(m,n)$ in (\ref{mfo}) can 
also be written as 
a non-local sum of Kogut-Susskind link electric fields 
along a line and the corresponding  parallel transports using (\ref{sunplaqe}). The magnetic charge  on the plaquette $p=(m,n)$ thus develops 
an infinite   Dirac string in the original (standard) $\{E^a(l); U(l)\}$ description. This is 
similar to the  discrete $Z_2$ disorder operator ${\Sigma}(m,n)$ written in terms of the original electric field operators $\sigma_1(m,n')$ in (\ref{z2do}) and (\ref{z2do2}). The Dirac string 
is shown in Figure \ref{mo}-b for our choice 
of canonical transformations. Note that it can be rotated by choosing a different scheme 
for the canonical transformations. Only the 
end points of  Dirac strings (location of the vortex)
are physical and independent of the canonical transformation schemes. This is exactly analogous to the rotations of the $Z_2$ Dirac strings  and the location of $Z_2$ vortices in (\ref{z2do}) or (\ref{z2do2}). Their orientations do not matter 
as Dirac strings themselves are invisible. In $Z_2$ or U(1) gauge theories they are 
trivially invisible as two horizontal links in Figure (\ref{mo})-b change by opposite phases which commute 
through the plaquette links and cancel each other. 
However, in the 
present non-abelian case, the non-local parallel transport $R_{ab}(S(m,n;n'))$ in relation (\ref{sunplaqe})  plays an important role in making the Dirac string invisible. These issues are 
further discussed in detail in Appendix {\ref{ids}}. 
We note that  the disorder operators in the strong coupling vacuum satisfy: 
$$\left\langle 0|\Sigma^{\pm}_\theta(m,n)|0\right\rangle =1.$$ 
The vacuum expectation 
values of all Wilson loops, on the other hand, are zero: $\left\langle 0|W_{C}(U)|0\right\rangle =0$.
It will be interesting to study the behavior of the 
vacuum expectation values of  
$\Sigma_\theta^\pm$ for the finite values of the coupling along with the vacuum correlation functions $\left\langle \Sigma^{\pm}_\theta(p)\Sigma^{\mp}_\theta(p')\right\rangle$, shown in Figure \ref{mo}-c, as $|p-p'| \rightarrow \infty$. The work in this direction is in progress and will be reported elsewhere.

\section{\bf A variational ground state of SU(N) spin model}
\label{vgs} 

In this section, we study the ground state of the 
dual spin model with nearest neighbor interactions.  
We then compare the results with those obtained from the variational analysis of the standard Kogut Susskind formulation \cite{arisue,suranyi,drstump}.
Note that 
after canonical transformations each plaquette loop is a fundamental  degree of freedom. Therefore gauge invariant computations in the dual spin model become much simpler. For simplicity 
we consider $N=2$. 
  For the ground state of SU(2) gauge theory, the magnetic fluctuations in a region are independent of fluctuation in another region sufficiently far away \cite{feynman,greensite}. So, the largest contributions to the vacuum state  comes from states with little magnetic correlations. Therefore we use the following separable state without any 
  spin-spin correlations as our variational ansatz: 
 \begin{align} 
   |\psi_0\rangle &= e^{S/2}  |0\rangle ; & S &= \alpha \sum\limits_{{p}} Tr{\cal W}({p}). \nonumber\\
   &=\prod\limits_p ~|\psi_0\rangle_p.
   \label{gswf} 
  \end{align} 
Above, $|0\rangle$ is the strong coupling vacuum state defined by ${\cal E}^a_\pm(m,n)|0\rangle=0$ and  $\alpha$ is  the variational parameter. 
Since this state doesn't have long distance correlations, it satisfies Wilson's area law criterion. 
We consider a Wilson loop $Tr~(W_C)$ along a large space loop C on the lattice and compute its ground state expectation value:
%
%
 $<\psi_0|Tr W_C|\psi_0>/~ \langle \psi_0|\psi_0\rangle$.
 \begin{figure}
  \includegraphics[scale=.65]{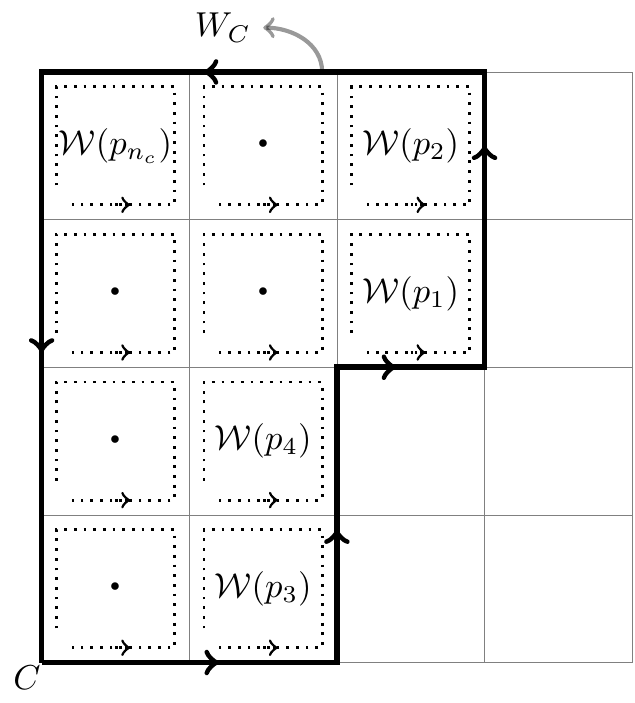}
  \caption{A Wilson loop $W_C$ can be written as the product of fundamental plaquette loop operators ${\cal W}(p)$. $W_C =  {\cal W}(p_1)~{\cal W}(p_2)~{\cal W}(p_3)\cdots\cdots {\cal W}(p_{n_c})$. The tails of the fundamental  plaquette loop operators connecting them to the origin 
  (see Figure \ref{dualsun}-a) are not shown for clarity.}
  \label{wlff}
  \end{figure}
As shown in the Appendix \ref{appb} (see (\ref{arealaw})):
\begin{align}
\hspace{-0.2cm} \frac{\langle\psi_0|Tr W_C|\psi_0\rangle}{\langle\psi_0|\psi_0\rangle} 
& = 2\left(\frac{I_2(2\alpha)}{I_1(2\alpha)}\right)^{n_c} 
 = 2e^{-n_c \ln \left(\frac{I_1(2\alpha)}{I_2(2\alpha)}\right)}
 \label{area1}
\end{align}
In (\ref{area1}), we have used the decomposition 
(\ref{dwl}) and $n_c$ is the number of plaquettes in the loop C. The function $I_l$ is the $l$-th order modified Bessel function of the first kind. The string tension is given by $\sigma_T(\alpha) = \ln \left(\frac{I_1(2\alpha)}{I_2(2\alpha)}\right)$.

We now calculate $\alpha$ by minimizing $$\langle H_{spin} \rangle = \frac{\langle\psi_0|H_{spin}|\psi_0\rangle}{\langle\psi_0|\psi_0\rangle}.$$
In order to calculate $\langle H_{spin} \rangle$, we first find the expectation value of ${\cal E}_-(p) \cdot {\cal E}_+(p')$ and ${\cal E}(p)\cdot {\cal E}(p)  \equiv   {\cal E}_+(p) \cdot {\cal E}_+(p) \equiv 
{\cal E}_-(p) \cdot {\cal E}_-(p)$ in (\ref{weakh}). This  calculation is 
done in Appendix \ref{appc}. The expectation values are (see (\ref{tevs})) 
$$\langle {\cal E}_-(p) \cdot {\cal E}_+(p') \rangle=0, ~~~~\langle {\cal E}(p) \cdot {\cal E}(p) \rangle = \frac{3\alpha}{16} \langle Tr{\cal W}(p) \rangle.$$
Putting $n_c=1$ in equation (\ref{area1}), we get $\langle Tr{\cal W}(p)\rangle=\frac{2 I_2(2\alpha)}{I_1(2\alpha)}$.
Therefore the expectation value of the effective Hamiltonian $H_{spin}$ is  
\begin{align}
\hspace{-0.2cm} \frac{\langle\psi_0|H_{spin}|\psi_0\rangle}{\langle\psi_0|\psi_0\rangle}
& \hspace{-0.1cm}= 2{\cal P} \hspace{-0.1cm} \left\{\left(\frac{3\alpha}{4}g^2-\frac{1}{g^2}\right)\frac{I_2(2\alpha)}{I_1(2\alpha)} +\frac{1}{g^2} \right\}.
\end{align}
Above,
 ${\cal P}$ is the number of plaquettes in the lattice.  $\frac{I_2(2\alpha)}{I_1(2\alpha)}$ is a monotonously increasing bounded function of $\alpha$. It takes values between $+1$ and $-1$ with $+1$ at $\alpha\rightarrow\infty$ and $-1$ at $\alpha\rightarrow-\infty$. In the weak coupling limit, $g^2\rightarrow 0$, $\frac{I_2(2\alpha)}{I_1(2\alpha)}$ should be maximum for the expectation value of $H_{spin}$ to be minimum and therefore, $\alpha \rightarrow \infty$.
   But, using the asymptotic form of the modified Bessel function of the first kind $I_l(2\alpha)$,
\[
I_l(2\alpha)  \xrightarrow{\alpha \rightarrow \infty} \frac{e^{2\alpha}}{\sqrt{2\pi(2\alpha)}}\left(1+\frac{(1-2l)(1+2l)}{16 \alpha}+\cdots\right) 
\]
In the weak coupling limit , $\frac{I_2(2\alpha)}{I_1(2\alpha)} \approx 1-\frac{3}{4\alpha}$.
Hence,
\begin{align}
\frac{\langle\psi_0|H_{spin}|\psi_0\rangle}{\langle\psi_0|\psi_0\rangle} &= \sum\limits_p 2 \left\{\left(\frac{3\alpha}{4}g^2-\frac{1}{g^2}\right)\left(1-\frac{3}{4\alpha}\right)+\frac{1}{g^2}\right\} 
\end{align}
Minimizing the expectation value in the weak coupling limit,
$\alpha=\frac{1}{g^2}$.
The string tension is given by $\sigma_T(\frac{1}{g^2})=ln\left(I_1(\frac{1}{g^2})/I_2(\frac{1}{g^2})\right)$.
This is
exactly the 
result obtained in \cite{suranyi,arisue} using  variational calculation with the fully disordered ground state and Kogut-Susskind Hamiltonian 
(\ref{hamks}) which is 
dual to the full non-local spin Hamiltonian. As shown in Appendix \ref{appc}, the expectation value of the non-local part of the Hamiltonian in the variational ground state $|\psi_0\rangle$ vanishes. So, 
 the simplified Hamiltonian with 
 nearest neighbor interactions
 gives the same variational ground state to the 
 lowest order  as the full Hamiltonian. 
 The disorder operator expectation value in this variational ground state is 
\begin{align}
 \frac{\langle\psi_0|\Sigma_{\theta=2\pi}^\pm(m,n)|\psi_0\rangle}{\langle\psi_0|\psi_0\rangle}  = \frac{16 \pi^2}{{}_{p}\langle\psi_0|\psi_0\rangle_{{}_{p}}}
= \frac{\pi\alpha}{2I_1(2\alpha)}. 
 \end{align}
 Above, we have defined $p\equiv (m,n)$ and written the separable state $|\psi_0\rangle$ as the direct product of the state vectors corresponding to each plaquette i.e, $|\psi_0\rangle= \prod_p~|\psi_0\rangle_p$. 

\section{Summary and Discussion}
In this work, we have shown that the 
canonical transformations provide a method to generalize Wegner duality between $Z_2$ lattice gauge theory and quantum Ising model to SU(N) lattice gauge theories. The similarities between $Z_2$ Wegner and SU(N) dualities were emphasized.   
The SU(N) dual formulation leads to 
a  new gauge invariant  disorder operator creating, annihilating magnetic vortices. This disorder operator  can be measured  in Monte-Carlo simulations. At $\theta =2\pi$ it reduces to 't Hooft disorder operator creating center vortices.
It will be interesting to see its behavior across the
deconfinement transition. 
 In the weak coupling continuum limit, the Hamiltonian of the dual model reduces to an effective SU(N) spin Hamiltonian with nearest neighbouring interactions.  We use a variational analysis of the spin model with a completely disordered ground state ansatz.  The effective spin Hamiltonian leads to the same results as the standard Kogut Susskind Hamiltonian.  Further analysis of the SU(N) spin model and its spectrum is required. This is the subject of our future 
 investigations.   It will also be interesting to generalize these transformations to $(3+1)$ dimensions to define dual electric vector potentials with a dual gauge group.  The work in this direction is also in progress.
 
\acknowledgments

\noindent {\it Acknowledgments: We thank  Ramesh Anishetty for many discussions throughout this work 
and for reading the manuscript.  
M.M would like to thank Michael Grady for  correspondence.  T.P.S thanks Council of Scientific and Industrial Research (CSIR), India for financial support.}

\appendix
\section{Wegner gauge-spin duality through canonical transformations}
\label{appa}

In this Appendix, we describe the canonical transformation  involved in the construction of the duality relation between the basic operators of $Z_2$ gauge theory and Ising model in $2+1$ dimensions. The net effect of the canonical transformation involved in the construction of the spin operators on a single plaquette, described in section \ref{sz2gtsm}, can be summarized as follows:
\begin{itemize}[leftmargin=*]
\item It replaces the top link $l_3$ on the plaquette by a plaquette spin operator with the same `electric field' as the top link:
{\footnotesize
\begin{align} \mu_{1}(p) &= \sigma_3(l_1)\sigma_3(l_2)\sigma_3(l_3)\sigma_3(l_4), & \mu_3(p) &= \sigma_1(l_3)\label{1pl1}.\end{align}
}
\item The `electric field' of the top link $l_3$ that vanishes gets absorbed into the electric fields of other links $l_1,l_2,l_4$:
{\footnotesize 
\begin{align} 
\bar{\sigma}_3(l_1)&=\sigma_3(l_1),&\bar{\sigma}_1(l_1)&=\sigma_1(l_1)\sigma_1(l_3)\nonumber\\
\bar{\sigma}_3(l_2)&=\sigma_3(l_2),&\bar{\sigma}_1(l_2)&=\sigma_1(l_2)\sigma_1(l_3)\nonumber\\
\bar{\sigma}_3(l_4)&=\sigma_3(l_4),&\bar{\sigma}_1(l_4)&=\sigma_1(l_4)\sigma_1(l_3)
\label{1pl2}
\end{align}
}
It is convenient to call the above net canonical transformation a `plaquette canonical transformation (C.T)'. We now generalize the duality transformation relation to a finite lattice by iterating the plaquette C.T all over the two dimensional lattice starting from the top left plaquette of the lattice and systematically repeating it from top to bottom and left to right.  
 We will illustrate this procedure on a $2 \times 2$ lattice which contains all the essential features of the construction on any finite lattice. The sites of the lattice are labelled as $O \equiv (0,0),A\equiv (0,1),B \equiv (0,2),C \equiv (1,0),D \equiv (1,1),E \equiv (1,2),F \equiv (2,0),G \equiv (2,1),H \equiv (2,2)$ and the plaquettes are numbered from top to bottom and left to right (see Figure \ref{22z2})  for convenience. The dual spin operators are constructed on a $2 \times 2$ lattice in 4 steps.
\end{itemize}   
\begin{figure*}
 \centering
  \includegraphics[scale=.67]{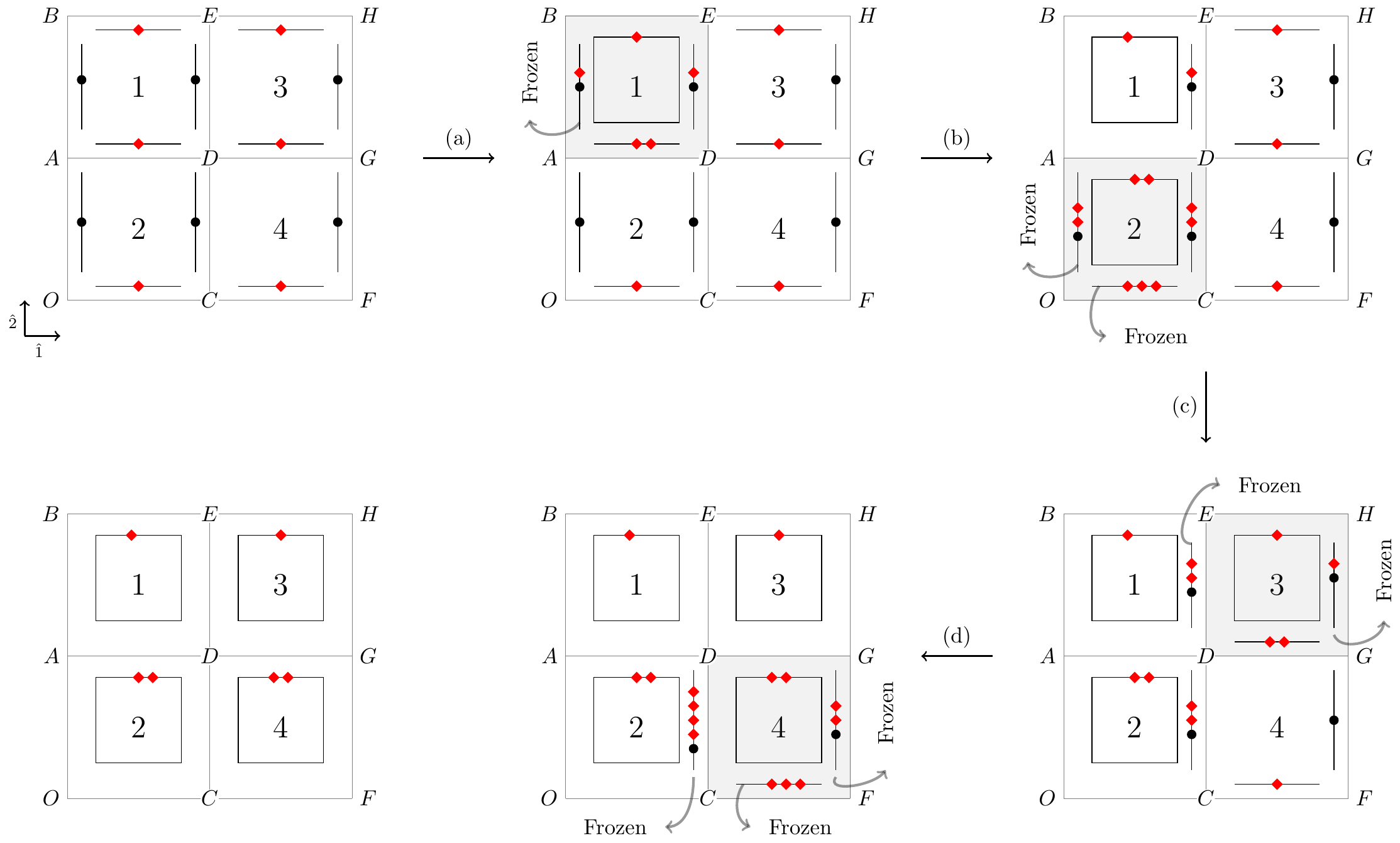}
  \caption{The `plaquette' canonical transformations involved in the construction of the duality transformation between $Z_2$ lattice gauge theory and $Z_2$ spin model on a $2 \times 2$ lattice. The steps (a), (b), (c) and (d) are plaquette canonical transformations on the plaquettes 1, 2, 3 and  4 respectively. 
  The electric field $\sigma_1(l)$ corresponding to the vertical and horizontal links are denoted by $\bullet$ and {\tiny {\color{red}$\blacklozenge$}} respectively. }
  \label{22z2}
 \end{figure*}
\begin{enumerate}[leftmargin=*]
\item 
We begin by performing the plaquette canonical transformation (\ref{1pl1}),(\ref{1pl2}) on plaquette 1. The spin conjugate operators $\{\mu_1(1);\mu_3(1)\}$ on plaquette 1 are  
{\footnotesize
\begin{align} 
&\mu_{1}(1)\equiv \mu_1(E) =\sigma_3(A;\hat{1})\sigma_3(D;\hat{2})\sigma_3(B;\hat{1})\sigma_3(A;\hat{2}), \nonumber\\ 
&~~~~~~~~~~~~~~~~\mu_3(1)\equiv \mu_3(E)  = \sigma_1(B;\hat{1}). 
\label{pl1spin}
\end{align}
}
The redefined link and string operators around plaquette 1 are
{\footnotesize
\begin{align} 
& \sigma_{3[x]}(D;\hat{2})=\sigma_3(D;\hat{2}), 
~~~\sigma_{1[x]}(D;\hat{2})=\sigma_1(D;\hat{2})\sigma_1(B;\hat{1}) 
\nonumber \\&
 \sigma_{3[x]}(A;\hat{2})=\sigma_3(A;\hat{2}),~~~
 \sigma_{1[x]}(A;\hat{2})=\sigma_1(A;\hat{2})\sigma_1(B;\hat{1}) \nonumber \\ &
 \sigma_{3[x]}(A;\hat{1})=\sigma_3(A;\hat{1}),~~~
 \sigma_{1[x]}(A;\hat{1})=\sigma_1(A;\hat{1})\sigma_1(B;\hat{1}). \nonumber 
\end{align}
}
Our notation is such that $\sigma_3(A;\hat{1})$ denotes the $\sigma_3$ variable of the link which starts at site A and is in the $\hat{1}$ direction. The subscript $[x]$ on $\sigma_{3[x]}(A;\hat{1})$ indicates that the electric field $\sigma_1(A;\hat{1})$ absorbs the electric field of the vanishing horizontal  link $(B;\hat{1})$ to become $\sigma_{1[x]}(A;\hat{1})$ during the plaquette C.T. Note that by our convention, the plaquette or spin operators are labelled by the top right corner of the plaquette. This plaquette C.T is illustrated in Figure \ref{22z2} (a). As a result of Gauss law at B: $$\sigma_{1[x]}(A;\hat{2}) \equiv {\cal G}(B) \approx 1.$$ Therefore, $\left\{ \sigma_{1[x]}(A;\hat{2}); \sigma_{3[x]}(A;\hat{2})
\right\} \equiv \left\{ \bar{\sigma}_{1}(B); \bar{\sigma}_{3}(B)\right\}$ are frozen and hence decouple from the physical Hilbert space. Again, as in the main text, the string operators are labelled by their right/top endpoints.   We are now left with the conjugate spin operators $\left\{ \mu_{1}(1); \mu_{3}(1)\right\}$ and the two link conjugate pair operators $\left\{ \sigma_{1[x]}(D;\hat{2}); \sigma_{3[x]}(D;\hat{2})\right\}, \left\{ \sigma_{1[x]}(A;\hat{1}); \sigma_{3[x]}(A;\hat{1})\right\}$. These link operators undergo further canonical transformations.  

\item We now iterate the plaquette C.T. on plaquette 2 to construct the spin or plaquette conjugate operators $\left\{\mu_{1}(2); \mu_{3}(2)\right\}$ and the link conjugate operators $\left\{ \sigma_{1[x]}(C;\hat{2});\sigma_{3[x]}(C;\hat{2})\right\}$, $\left\{ \sigma_{1[x]}(O;\hat{1}); \sigma_{3[x]}(O;\hat{1})\right\}$, $\left\{ \sigma_{1[x]}(O;\hat{2}); \sigma_{3[x]}(O;\hat{2})\right\}$ as illustrated in Figure \ref{22z2}-b. The spin operators are
{\footnotesize
\begin{align}
&\mu_{1}(2)\equiv \mu_1(D)=\sigma_{3}(A;\hat{1})\sigma_3(O;\hat{2})\sigma_3(O;\hat{1})\sigma_3(C;\hat{2}), \nonumber\\
& \mu_3(2)\equiv \mu_3(D)=\sigma_{1[x]}(A;\hat{1})=\sigma_1(A;\hat{1})\sigma_1(B;\hat{1}) 
\end{align}
} 
The redefined link and new string operators around plaquette 2 are
{\footnotesize
\begin{align} 
& \sigma_{3[x]}(C;\hat{2})=\sigma_3(C;\hat{2}), \nonumber\\& \sigma_{1[x]}(C;\hat{2})=\sigma_1(C;\hat{2})\sigma_{1[x]}(A;\hat{1})= \sigma_1(C;\hat{2})\sigma_1(A;\hat{1})\sigma_1(B;\hat{1}) \nonumber  \\
&\sigma_{3[x]}(O;\hat{1})=\sigma_3(O;\hat{1}), \nonumber
\end{align}
}
{\footnotesize
\begin{align}
& \sigma_{1[x]}(O;\hat{1})=\sigma_1(O;\hat{1})\sigma_{1[x]}(A;\hat{1})= {\cal G}(O){\cal G}(A){\cal G}(B)\approx 1 \nonumber\\
&\sigma_{3[x]}(O;\hat{2})=\sigma_3(O;\hat{2}), \\& \sigma_{1[x]}(O;\hat{2})=\sigma_1(O;\hat{2})\sigma_{1[x]}(A;\hat{1})= {\cal G}(A){\cal G}(B) \approx 1. \nonumber
\end{align}
}
Thus the string conjugate pairs  $\left\{\sigma_{1[x]}(O;\hat{1});\sigma_{3[x]}(O;\hat{1})\right\} \equiv \left\{\bar{\sigma}_1(C); \bar{\sigma}_3(C)\right\}$ and $\left\{ \sigma_{1[x]}(O;\hat{2});\sigma_{3[x]}(O;\hat{2})\right\}\equiv \left\{\bar{\sigma}_1(A); \bar{\sigma}_3(A)\right\}$ are frozen due to Gauss law at O, A and B. 

\item The third step involves iterating the plaquette C.T. on plaquette 3 as shown in Figure \ref{22z2}(c). This leads to decoupling of $\left\{ \sigma_{1[x]}(G;\hat{2});\sigma_{3[x]}(G;\hat{2})\right\} \equiv \left\{\bar{\sigma}_1(H);\bar{\sigma}_3(H)\right\}$, $\left\{ \sigma_{1[xx]}(D,\hat{2});\sigma_{3[xx]}(D,\hat{2})
\right\}  \equiv \left\{\bar{\sigma}_1(E);\bar{\sigma}_3(E)\right\}$ due to the $Z_2$ Gauss laws at E and H. The canonical transformations 
on plaquette 3 defining the spins are
\begin{align} 
&\mu_{1}(3)\equiv \mu_1(H) = \sigma_3(E;\hat{1})\sigma_3(G;\hat{2})\sigma_3(D;\hat{1})\sigma_3(D;\hat{2}), \nonumber\\
& \mu_3(3)\equiv \mu_3(H) =\sigma_3(E;\hat{1}). 
\end{align} 
The redefined links and strings around plaquette 3 are
\begin{align} 
&  \sigma_{3[xx]}(D,\hat{2})=
\sigma_{3[x]}(D;\hat{2})=\sigma_3(D;\hat{2}), \nonumber\\& \sigma_{1[xx]}(D;\hat{2})=\sigma_{1[x]}(D;\hat{2})\sigma_1(E;\hat{1})= {\cal G}(E)\approx 1\nonumber \\& 
\sigma_{3[x]}(G;\hat{2})=\sigma_3(G;\hat{2}), \nonumber\\& \sigma_{1[x]}(G;\hat{2})=\sigma_1(G;\hat{2})\sigma_1(E;\hat{1}) = {\cal G}(H)\approx 1 \nonumber \\
& \sigma_{3[x]}(D;\hat{1})=\sigma_3(D;\hat{1}), \nonumber\\
& \sigma_{1[x]}(D;\hat{1})= \sigma_1(D;\hat{1})\sigma_1(E;\hat{1})
\end{align}
\item Finally, we iterate the plaquette C.T. on plaquette 4  which are shown in Figure \ref{22z2}(d).
The conjugate spin operators $\{\mu_1(4);\mu_3(4)\}$ on plaquette 4 are 
{\footnotesize
\begin{align}
&\mu_{1}(4)\equiv \mu_1(G) = \sigma_3(D;\hat{1})\sigma_3(F;\hat{2})\sigma_3(C;\hat{1})\sigma_3(C;\hat{2}),\nonumber\\
& \mu_3(4)\equiv \mu_3(G) = \sigma_{1[x]}(D;\hat{1})=\sigma_1(D;\hat{1})\sigma_1(E;\hat{1})
\end{align}
} 
The remaining string operators are
{\footnotesize
\begin{align}
&\sigma_{3[xx]}(C;\hat{2})= \sigma_{3[x]}(C;\hat{2})=\sigma_3(C;\hat{2}),\nonumber\\& \sigma_{1[xx]}(C;\hat{2})=\sigma_{1[x]}(C;\hat{2})\sigma_{1[x]}(D;\hat{1})={\cal G}(D){\cal G}(E) \approx 1\nonumber \\&
\sigma_{3[x]}(C;\hat{1})=\sigma_3(C;\hat{1})\nonumber\\& \sigma_{1[x]}(C;\hat{1})=\sigma_1(C;\hat{1})\sigma_{1[x]}(D;\hat{1})\nonumber\\& 
\hspace{1.5cm} = {\cal G}(C){\cal G}(O){\cal G}(A){\cal G}(D){\cal G}(B){\cal G}(E) \approx 1 \nonumber\\
&\sigma_{3[x]}(F;\hat{2})=\sigma_3(F;\hat{2}), \\& \sigma_{1[x]}(F;\hat{2})= \sigma_1(F;\hat{2})\sigma_{1[x]}(D;\hat{1})={\cal G}(G){\cal G}(H)\approx 1. \nonumber 
\end{align}
}
Gauss laws at $O,A,B,C,D,E,G$ and H implies that the remaining string operators 
{\footnotesize{$\left\{ \sigma_{1[xx]}(C;\hat{2});\sigma_{3[xx]}(C;\hat{2})\right\}  \equiv \left\{\bar{\sigma}_1(D);\bar{\sigma}_3(D)\right\}$, $\left\{ \sigma_{1[x]}(C,\hat{1});\sigma_{3[x]}(C,\hat{1})
\right\}  \equiv \left\{\bar{\sigma}_1(F);\bar{\sigma}_3(F)\right\}$ and $\left\{ \sigma_{1[x]}(F,\hat{2});\sigma_{3[x]}(F,\hat{2})\right\}  \equiv \left\{\bar{\sigma}_1(G);\bar{\sigma}_3(G)\right\}$}} are frozen. As a result, after the series of 4 plaquette C.T.s, all the Gauss law constraints are solved. Only the plaquette/spin variables $\left\{\mu_1(1);\mu_3(1) \right\}, \left\{\mu_1(2);\mu_3(2) \right\}, \left\{\mu_1(3);\mu_3(3) \right\}$ and $\left\{\mu_1(4);\mu_3(4) \right\}$ remains in the physical Hilbert space. This leads to a dual $Z_2$ spin model.
These results can be directly generalized to any finite lattice without any new issues, to give the duality relations (\ref{plaqop1}),(\ref{plaqop2}), (\ref{z2duality1})-(\ref{z2duality2}).  
\end{enumerate}
\section{Non-abelian Duality \&  Non-locality}
\label{vortex}
In this Appendix we discuss the role of non local terms in non-abelian duality transformations. We show that 
they lead to magical cancellations required for (a) the SU(N) Gauss laws to be satisfied, (b) the Dirac strings associated with U(1) vortices to be invisible.    
These non-trivial cancellations  should be contrasted 
 with the similar but trivial $Z_2$ lattice gauge theory cancellations in terms of the electric scalar potentials discussed in section \ref{sz2glt} and illustrated  in Figure \ref{z2glc}.
\subsection{SU(N) Gauss laws \& Solutions}
\label{nagls} 

In this part we show that the dual SU(N) electric scalar potentials solve the local SU(N) Gauss laws at all lattice sites except the origin.  
At the origin 
they lead to the global constraints (\ref{resgl}).
This is SU(N) generalization of the $Z_2$ results 
discussed in section \ref{sz2glt}.
 
We show explicit calculations for all the possible cases namely ($m\neq0,n\neq0$),
($m=0,n\neq 0$), ($m\neq 0,n=0$) and ($m=0,n=0$).
\begin{widetext} 
 	\begin{itemize}
		\item $m\neq0,n\neq0$
		 
		The Kogut-Susskind SU(N) electric fields meeting at $(m,n)$  in terms of the  
		spin operators (\ref{kstow}) are:
		{\footnotesize
			\begin{align}
			&\hspace{-.2cm} E_+^a(m,n;\hat{1})= \underbrace{R_{ab}({\sf T}^\dagger(m,n))\bigg[{\cal E}^b_+(m+1,n)+{\cal E}_-^b(m+1,n+1)\bigg]}_{T_1} \nonumber \\
			&\hspace{-.2cm}E_+^a(m,n; \hat{2})= \underbrace{R_{ab}({\sf T}^\dagger(m,n))\bigg[\sum\limits_{\bar{n}=n+2}^{{{\sf N}}} \{{\cal E}_+^b(m+1,\bar{n})+{\cal E}_-^b(m+1,\bar{n})\}+{\cal E}_+^b(m+1,n+1)\bigg]}_{T_2} + 
			\underbrace{R_{ab}(U^\dagger(m-1,n;\hat{1})){\sf T}^\dagger(m-1,n) {\cal E}_-^b(m,n+1)}_{T_3}\nonumber \\
			&\hspace{-.2cm}E_-^a(m,n;\hat{1})=\underbrace{-R_{ab}({U}^\dagger(m-1,n;\hat{1}){\sf T}^\dagger(m-1,n){\cal E}_-^b(m,n+1)}_{-T_3}+ \underbrace{-R_{ab}({U}^\dagger(m-1,n;\hat{1}){\sf T}^\dagger(m-1,n)){\cal E}_+^b(m,n)}_{-T_4} \nonumber \\
			&\hspace{-.2cm}E_-^a(m,n;\hat{2})= \underbrace{-R_{ab}({\sf T}^\dagger(m,n))\bigg\{{\cal E}_+^b(m+1,n) + {\cal E}_-^b(m+1,n+1)\bigg\}}_{-T_1}+
			\\ 
			& \hspace{1.4cm}\underbrace{-R_{ab}({\sf T}^\dagger(m,n))\bigg\{{\cal E}_+^b(m+1,n+1) +\sum\limits_{\bar{n}=n+2}^{{\sf N}}\left[{\cal E}_+^b(m+1,\bar{n})+{\cal E}_-^b(m+1,\bar{n})\right]\bigg\}}_{-T_2} +\underbrace{R_{ab}(U^\dagger(m-1,n,\hat{1}){\sf T}^\dagger(m-1,n)) {\cal E}_+^b(m,n)}_{T_4}.\nonumber
			\label{redundantgl}
			\end{align}}
Therefore all $T_1,T_2,T_3,T_4$ cancel and  ${\cal G}^a(m \neq 0,n \neq 0)=E_+^a(m,n;\hat{1})+E_+^a(m,n;\hat{2})+E_-^a(m,n;\hat{1})+E_-^a(m,n;\hat{2}) 
		\equiv 0.$ 
		
		\item $m=0,~n\neq0$
		
		Similarly, the SU(N) electric fields at  $(0,n\neq0)$ in terms of dual  SU(N) spin operators in (\ref{kstow}) are
		\begin{align}
		& E_+^a(0,n;\hat{1})=R_{ab}({\sf T}(0,n))\big\{\underbrace{{\cal E}^b_-(1,n+1)}_{T_1} + \underbrace{{\cal E}^b_+(1,n)}_{T_2} \big\} \nonumber\\
		&E_+^a(0,n;\hat{2})=R_{ab}({\sf T}(0,n))\big\{\underbrace{{\cal E}^b_+(1,n+1)}_{T_3}+\underbrace{\sum\limits_{\bar{n}=n+2}^{\sf N} {\mathbb L}^b(1,\bar{n})}_{T_4}\big\} \\
		&E_-^a(0,n;\hat{2})= -R_{ab}(U(0,m,\hat{2}))E_+^b(0,n-1,\hat{2})=
		-R_{ab}({\sf T}(0,n))\big\{{\cal E}^b_+(1,n)+\sum\limits_{\bar{n}=n+1}^{\sf N}{\mathbb L}^b(1,\bar{n})\big\} \nonumber \\
		&= -R_{ab}({\sf T}(0,n))\big\{{\cal E}^b_+(1,n)+ {\mathbb L}^b(1,{n+1})+\sum\limits _{\bar{n}=n+2}^{\sf N}{\mathbb L}^b(1,\bar{n})\big\} \nonumber \\
		& = -R_{ab}({\sf T}(0,n))\big\{\underbrace{{\cal E}^b_+(1,n)}_{T_2}+ \underbrace{{\cal E}^b_+(1,n+1)}_{T_3}+
		\underbrace{{\cal E}^b_-(1,n+1)}_{T_1} 
		 +\underbrace{\sum\limits _{\bar{n}=n+2}^{\sf N}{\mathbb L}^b(1,\bar{n})}_{T_4}\big\}.
		\end{align}
		All $T_1,T_2,T_3,T_4$ cancel and 
		 ${{\cal G}^a(0,n \neq 0)}
		=E_+^a(0,n;\hat{1})+E_+^a(0,n;\hat{2})+E_-^a(0,n;\hat{2}) \equiv 0$.   
		
		\item $m\neq0,n=0$
		
		The Kogut-Susskind electric fields at site $(m\neq0,0)$ are 
		\begin{align}
		& E_+^a(m,0,\hat{1})=R_{ab}({\sf T}(m,0))\Big\{ \underbrace{{\cal E}_-^b(m+1,1)}_{T_1}+\underbrace{\sum\limits_{\bar{m}=m+2}^{\sf N} \sum\limits_{\bar{n}=1}^{\sf N} {\mathbb L}^b(\bar{m},\bar{n})}_{T_2}\Big\}\nonumber\\
		& E_+^a(m,0,\hat{2})=R_{ab}({\sf T}(m,0))\Big\{ \underbrace{{\cal E}_+^b(m+1,1)}_{T_3}+\underbrace{{\cal E}_-^b(m,1)}_{T_4}+\underbrace{\sum_{\bar{n}=2}^{\sf N} {\mathbb L}^b(m+1,\bar{n})}_{T_5}\Big\}\\
		& E_-^a(m,0,\hat{1})=R_{ab}({\sf T}(m-1,0,\hat{1}))E_+^a(m-1,0,\hat{1})=-R_{ab}({\sf T}(m,0)) \Big\{{\cal E}_-^b(m,1)+\sum\limits_{\bar{m}=m+1}^{\sf N}\sum\limits_{\bar{n}=1}^{\sf N} {\mathbb L}^b(\bar{m},\bar{n})\Big\}\nonumber \\ 
		& =-R_{ab}({\sf T}(m,0)) \Big\{\underbrace{{\cal E}_-^b(m,1)}_{T_4}+ \underbrace{{\cal E}^b_+(m+1,1)}_{T_3} +\underbrace{{\cal E}^b_-(m+1,1)}_{T_1}+ 
		\underbrace{\sum\limits_{\bar{n}=2}^{\sf N} {\mathbb L}^b(m+1,\bar{n})}_{T_5}
		+\underbrace{\sum\limits_{\bar{m}=m+2}^{\sf N}\sum\limits_{\bar{n}=1}^{\sf N} {\mathbb L}^b(\bar{m},\bar{n})}_{T_2}\Big\} 
		\end{align} 
		As before, all $T_1,T_2,T_3,T_4,T_5$ cancel leading to ${\cal G}^a(m\neq 0,n=0)=E_+^a(m,0;\hat{1})+E_+^a(m,0;\hat{2})+E_-^a(m,0;\hat{1}) \equiv 0$. 
		
		\item $m=0,~n=0$
		
		The electric fields $E_+^a(0,0;\hat{1})$ and $E_+^a(0,0;\hat{2})$ in (\ref{kstow})
		are
		\begin{align}
		\hspace{2cm} E_+^a(0,0;\hat{1})={\cal E}_-^a(1,1)+\sum\limits_{\bar{m}=2}^{{\sf N}}\sum\limits_{\bar{n}=1}^{{\sf N}} {\mathbb L}^a(\bar{m},\bar{n})
		\hspace{2cm}E_+^a(0,0;\hat{2})={\cal E}_+^a(1,1)+\sum\limits_{\bar{n}=2}^{{\sf N}} {\mathbb L}^a(1,\bar{n}).
		\end{align}
		Therefore the Gauss law operator at the origin is given by:
		\begin{align}
		{\cal G}^a(0,0)= E_+^a(0,0;\hat{1})+E_+^a(0,0;\hat{2})=\sum\limits_{\bar{m}=1}^{{\sf N}}\sum\limits_{\bar{n}=1}^{{\sf N}} {\mathbb L}^a(\bar{m},\bar{n}) =0.
		\label{glp}
		\end{align}
	\end{itemize}
\end{widetext}
Thus the $SU(N)$ Gauss law constraints at the origin are not redundant and  lead to the residual global SU(N) Gauss law (\ref{resgl}) in terms of the SU(N) 
spin operators.  
Note  that in the abelian $U(1)$ case these cancellations are trivial as there are no color indices and $R_{ab}(U)\rightarrow 1$.  
The global Gauss law constraint identically vanishes as ${\mathbb L} \equiv 0$ in the abelian case. 

\subsection{The invisible Dirac strings}
\label{ids} 

\begin{figure*}[t]
			\includegraphics[]{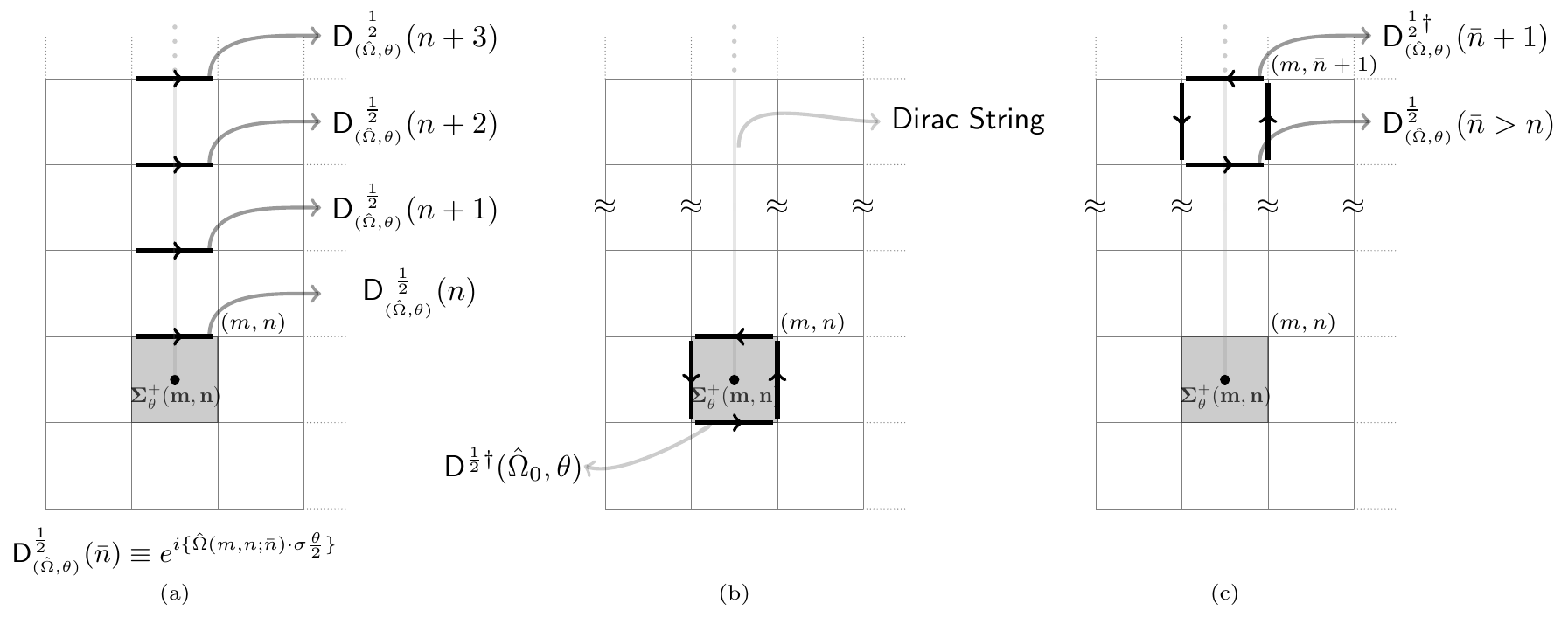}
			\caption{The invisible Dirac strings. The disorder operator $\Sigma_\theta^+(m,n)$ rotates (a) all  horizontal link operators $U(m-1,\bar{n}\geq n;\hat{1})$ by a factor ${\sf D}^{\frac{1}{2}}_\theta(\bar{n}) \equiv D^{\frac{1}{2}}(\hat \Omega(m,n;\bar n),\theta)$
			 in (\ref{omg}), (\ref{wm}), (b) the plaquette operator ${U}^p(m,n)$ rotates by  ${\sf D}^{\frac{1}{2}\dagger}(\hat \Omega_0,\theta)$ as in (\ref{su2disa1}) (c) all other plaquette  operators ${U}^p(m,\bar{n}> n)$  do not rotate. As shown in (\ref{su2disa2}), the two factors ${\sf D}^{\frac{1}{2}\dagger}_\theta(\bar n )$ and ${\sf D}^{\frac{1}{2}}({\bar n}+1)$ cancel highlighting 
			the role of non-locality in non-abelian duality. The  plaquette loop operators $U^p(\bar m\neq m, \bar n)$ are trivially invariant under $\Sigma_\theta^+(m,n)$. At
			$\theta =2 \pi$,  all ${\sf D}_{(\hat \Omega,\theta=2\pi)}^{~\frac{1}{2}}(n) =-1$
			in (a) and  all 
			dark horizontal links change sign creating a  $Z_2$ center vortex at $(m,n)$ \cite{hooft}.
			}
				\label{su2disfig}
		\end{figure*}
In this part we show how the SU(N) duality transformations make the non-local Dirac 
strings invisible.
We start with the  Dirac strings in the 
simple $Z_2$ lattice gauge theory. 
The $Z_2$ magnetic vortex operators (\ref{z2do2}) 
is 
\begin{eqnarray}
 {\Sigma}_{\pi}(m,n)
 =  exp ~i   \Bigg\{\sum_{\bar n=n}^{\infty} E(m-1,\bar n; \hat 1) ~\pi \Big\}. 
\label{z2do22} 
\end{eqnarray}
It is clear that this rotation operator by $\pi$ 
flips all 
$$\sigma_3(m',\bar n;\hat 1),~~~~m'=m-1, ~\bar n \ge n$$ 
along an infinitely long 
vertical Dirac string. The remaining operators 
over the entire lattice remain unaffected.  
Note that only the end point of the Dirac string is 
visible where $\Sigma_{\pi}(m,n)$ creates 
a magnetic vortex at $(m,n)$. We now 
generalize this result to SU(2) lattice gauge theory.
The SU(2) disorder operator is  
\begin{align}
&~~~{\Sigma }^{+}_\theta(m,n) \equiv exp ~i  \Bigg\{\left({\hat w }(m,n)\cdot {\cal E}_{+}(m,n)\right) \theta \Bigg\}\nonumber  \\ 
& ~=exp ~i~ 
\Bigg\{
~{\tiny{\sum\limits_{\bar n=n}^{\infty}}} 
\hat \Omega(m,n;\bar n) \cdot E_-(m,\bar n;\hat{1})~\theta\Bigg\}.
\label{mfoo} 
\end{align} 
In (\ref{mfoo}) we have used the defining equation (\ref{sunplaqe}) for the electric scalar potentials
\begin{align}
     {\mathcal E}_+^a(m,n) = \sum\limits_{\bar n=n}^{\sf N} R_{ab}(S(m,n;\bar n))~E_-^b(m,\bar n; \hat{1}). \nonumber  
\end{align} 
and 
\begin{align} 
\hat \Omega^a(m,n;\bar n) \equiv  \hat w^b(m,n)~R_{ba}(S(m,n;\bar n)).
\label{omg}
\end{align}  
From (\ref{mfoo}) it is clear that the SU(2) disorder operator $\Sigma_\theta^+(m,n)$ rotates only 
the following Kogut-Susskind flux  operators: 
\begin{align} 
U({\bar m},\bar {n};\hat{1}); ~~~~  {\bar m}=m-1, ~\bar {n} \ge n.
\label{efl}
\end{align}
 All other link operators $U(\bar m, \bar n,\hat i)$ over the entire lattice remain unaffected.  
 We now  show that the disorder operator $\Sigma_\theta^+(m,n)$ rotates only ${U}_p(m,n)$
  and leaves all other plaquette operators ${U}_p(\bar {m},\bar {n});~(\bar {m},\bar {n})\neq (m,n)$ unaffected.   The  non-local parallel transport operators $R_{ab}(S(m,n;\bar n))$ in non-abelian duality transformations play extremely crucial role 
in the cancellations involved. 

We consider two relevant cases: (a) The plaquette 
$U_p(m,n)$, ~~(b) The plaquettes $U_p(m,\bar n > n)$.
The action of  the disorder operator on these
plaquettes is graphically illustrated in Figure \ref{su2disfig}. Any plaquette $U^p(\bar m,\bar n)$ with 
$\bar m\neq m$ is not affected as none of the affected links (\ref{efl}) are present. 
\begin{itemize}[leftmargin=*]
		\item The plaquette $U_p(m,n)$: 
		
		This case is illustrated in Figure \ref{su2disfig}-b. For convenience sake, we define $U^p(m,n)=
		U(m-1, n-1;\hat 1) ~U(m, n-1;\hat 2)~ U^\dagger(m-1, n;\hat 1) ~U^\dagger(m-1, n-1; \hat 2) \equiv U_1~U_2~U_3^\dagger ~U_4^\dagger.$ From (\ref{mfoo}) it is clear that the disorder
		operators $\Sigma^{\pm}_\theta(m,n)$ act only on $U^\dagger(m-1, n,\hat 2) \equiv U^\dagger_3$ resulting in the rotation of the $U^p(m,n)$:  
		\begin{widetext}   
		{\footnotesize
			\begin{align}
			&\hspace{-.2cm}\Sigma_\theta^+(m,n){U}^p_{\alpha\beta}(m,n)\Sigma_\theta^-(m,n)
			=\Big(U_1~U_2\Big)_{\alpha\gamma}
			\Bigg\{\Sigma_\theta^+(m,n)~\left(U^\dagger_3\right)_{\gamma\delta}~\Sigma_\theta^-(m,n)
			\Bigg\}
			\left(U_4\right)_{\delta\beta}  
		= \Big(U_1~U_2\Big)_{\alpha\gamma}~
			\left(~D^{\frac{1}{2}\dagger} \Big(\hat \Omega(m,n;n),\theta \Big)~ 
			U^\dagger_3 \right)_{\gamma\delta}~       
			\left(U^\dagger_4\right)_{\delta\beta} 
			\nonumber\\
			&
			~~= \Bigg\{\left(U_1~U_2\right) ~D^{\frac{1}{2}\dagger} \Big(\hat \Omega(m,n;n),\theta \Big)~\left(U_1U_2\right)^\dagger 
			\Bigg\}_{\alpha\gamma} ~U^p_{\gamma\beta}(m,n) 
			= \Bigg(D^{ \frac{1}{2}\dagger} 
			\Big(\hat \Omega_0(m,n),\theta \Big)\Bigg)_{\alpha\gamma}~ U^p_{\gamma \beta}(m,n) \equiv {\sf D}_{\alpha\gamma}^{\frac{1}{2}}(\hat \Omega_0,\theta)U_{\gamma\beta}^p(m,n).
			\label{su2disa1}
			\end{align}
		}	
		\end{widetext}
		In (\ref{su2disa1})  $  
		D^{\frac{1}{2}}\Big(\hat \omega,\theta\Big)$ are the Wigner rotation matrices implementing rotations by  $\theta$ around $\hat \omega$ axis in the spin half representation: 
		\begin{align} 
		D^{\frac{1}{2}}(\hat \omega,\theta) \equiv exp~i \left(\hat \omega \cdot \sigma\right) \frac{\theta}{2}
		\label{wm}
		\end{align} 
		In the first step we have used the canonical commutation relations (\ref{ccr11}) to get these 
		Wigner matrices. In the third step 
		we have used:  
		$U~{\sigma^a}~ U^\dagger = R_{ab}(U^\dagger) ~\sigma^b$		
with $U \equiv U_1U_2$
and $R_{ab}(U)$ defined in (\ref{lrefr}). Using 
(\ref{omg}) and the above relations, the final 
axis of rotation in (\ref{su2disa1}) is
\begin{align} 
&\hat \Omega^a_0(m,n) 
= \hat \omega^b(m,n)R_{ba}\left({\sf T}(m-1,n-1)U_p(m,n)\right).
\label{omega2} 
\end{align}
In (\ref{omega2}) the parallel transport string ${\sf T}(m-1,n-1)$ is defined in (\ref{sunstring}). It is clear from (\ref{su2disa1}) that under the action of the disorder operator $\Sigma_\theta^+(m,n)$, the SU(N) spin operators ${\cal W}_{\alpha\beta}(m,n)$ gets rotated to $D^{\frac{1}{2}}_{\alpha\gamma}(\hat{w},\theta){\cal W}_{\gamma\beta}(m,n)$. Therefore, (\ref{su2disa1}) is consistent with (\ref{aabbcc}).

\item The plaquettes $U^p({m},\bar{n}>n)$ 
		
		This case is illustrated in Figure \ref{su2disfig}-c.
		 Again for convenience we write $U^p(m,\bar n>n) 
		= U_1U_2U^\dagger_3U^\dagger_4 \equiv U(m-1,\bar n-1,\hat1) U(m,\bar n-1,\hat 2) U^\dagger(m-1,\bar n,\hat 2) U^\dagger(m-1,\bar n-1,\hat 2).$
		\begin{widetext} 
		{\footnotesize
			\begin{align}
			&\Sigma_\theta^-(m,n)~{U}^{p}(m,\bar{n}>n)~\Sigma_\theta^+(m,n) = \Bigg\{\Sigma_\theta^-(m,n)~U_1~\Sigma_\theta^+(m,n)\Bigg\}U_2 
			\Bigg\{\Sigma_\theta^-(m,n)~ U^\dagger_3~ \Sigma_\theta^+(m,n) \Bigg\}~U^\dagger_4 = 
			\Bigg\{
			U_1~D^{\frac{1}{2}}\left(\hat \Omega(m,n; \bar n),\theta\right)\Bigg\}~U_2\nonumber \\& \Bigg\{ D^{\frac{1}{2}\dagger}\left(\hat \Omega(m,n;\bar n+1),\theta\right) U^\dagger_3\Bigg\}~U_4^{\dagger}
			 \equiv U_1D^{\frac{1}{2}}\left(\hat \Omega(m,n;\bar n),\theta\right) \underbrace{\Bigg\{U_2~ D^{\frac{1}{2}\dagger}\left(\hat \Omega(m,n;\bar n+1),\theta\right) U_2^\dagger\Bigg\}}_{D^{\frac{1}{2}\dagger}\left(\hat \Omega(m,n;\bar n),\theta\right)} U_2U_3^\dagger U_4^{\dagger} = U_1U_2U_3^\dagger U_4^\dagger =U^p(m,\bar n).
			\label{su2disa2}
			\end{align}
		}	
		\end{widetext} 
		\end{itemize}
In the last step we have used the following simple result: 
\footnotesize{
\begin{align} 
D^{\frac{1}{2}}(\hat \Omega(m,n;\bar n),\theta) = U(m,\bar n,\hat 2)~ D^{\frac{1}{2}}(\hat \Omega(m,n;\bar n+1),\theta)
  U^\dagger(m,\bar n,\hat 2)
  \label{form} 
  \end{align} 
  }
  Thus we see that the infinite number of  cancellations crucially depends on the very specific  form of the non-local parallel transport operators $R_{ab}(m,n;n')$ in (\ref{sunplaqe}) involved in the non-abelian duality.

\section{The ground state \& area law} 
\label{appb} 

In this Appendix, we calculate the expectation value of a large Wilson loop $W_C$ in the variational ground state $|\psi_0\rangle$ and show that it satisfies Wilson's Area law criterion. Any Wilson loop can be written as the product of fundamental plaquette loop operators, $W_C= \prod\limits_{p_i} {\cal W}(p_i)$. Here, $p_i$ denotes the plaquettes inside the loop $C$ in the order bottom right to top left (See Figure \ref{wlff}).  
It is convenient to define a complete basis $\prod_p |\omega_p,\hat{w}_p\rangle$,which diagonalises all Wilson loops. Above, $\prod_p$ is over all the plaquettes in the lattice and
\begin{align}
\hspace{-0.2cm}\ket{\omega_p,\hat{w}_p}=
\sum\limits_{jm_- m_+} \sqrt{2j+1}~ D^{~~j}_{m_-m_+}(\hat{w}_p,\omega_p) \ket{jm_- m_+}_p.
\label{angbasis}
\end{align}
In (\ref{angbasis}), $D^{~~j}_{m_-m_+}(\hat{w}_p,\omega_p)$ is a Wigner D matrix  in spin $j$ representation and $|j m_- m_+\rangle_p$ is the eigenbasis of $\left(\vec{\cal E}_+(p)\right)^2=\left(\vec{\cal E}_-(p)\right)^2 \equiv \left(\vec{\cal E}(p)\right)^2$, $\vec{\cal E}_-^{a=3}(p)$ and $\vec{\cal E}_+^{a=3}(p)$. Also, $(\omega_p,\hat{w}_p)$ is the angle axis parameterization of the $SU(2)$ group element associated with the plaquette loop operator at $p$, as defined in section \ref{sdo}.
The plaquette loop operator ${\cal W}_{\alpha \beta}(p)$ is diagonal in this basis, 
\[ {\cal W}_{\alpha \beta}(p)|\omega_p,\hat{w}_p\rangle = z_{\alpha\beta}(p)|\omega_p,\hat{w}_p \rangle .\]
with  
\[z_{\alpha \beta}=\begin{bmatrix}\left( \cos\frac{\omega}{2}-i\sin\frac{\omega}{2}\cos\theta \right)&i\sin\frac{\omega}{2}\sin\theta e^{-i\phi}\\-i\sin\frac{\omega}{2}\sin\theta e^{-i\phi}&\left( \cos\frac{\omega}{2}+i\sin\frac{\omega}{2}\cos\theta \right) \end{bmatrix}_{\alpha \beta}\]
where $\theta$ and $\phi$ are the angles characterizing $\hat{w}_p$. In particular,
\[Tr{\cal W}(p) |\omega_p,\hat{w}_p\rangle = 2\cos(\omega_p/2) |\omega_p,\hat{w}_p\rangle. \]
The expectation value of $TrW_C$ in $|\psi_0\rangle$ is given by   
{\footnotesize  
\begin{align}
&\langle TrW_C\rangle~\equiv~\frac{\langle\psi_0|TrW_C|\psi_0\rangle}{\langle\psi_0|\psi_0\rangle} \nonumber\\
&= \frac{1}{\langle\psi_0|\psi_0\rangle}\prod\limits_{p\in p_i} \int d\mu(\omega_p,\hat{w}_p)\big\langle0|e^S Trz(C)|\omega_p,\hat{w}_p\big\rangle \big\langle\omega_p,\hat{w}_p|0\big\rangle\nonumber\\
&= \frac{\prod_{p}  
\int d\mu(\omega_p,\hat{w}_p)~e^{2\alpha \cos{\omega_p/2}} ~2\cos\left(\omega(C)/2\right)}{\prod_{p}   
\int d\mu(\omega_p,\hat{w}_p)~e^{2\alpha \cos(\frac{\omega_{p}}{2})}  } 
\label{area}
\end{align}
}
In (\ref{area}),  
$\int d\mu(\omega_p,\hat{\omega}_p)\equiv \int\limits_0^{2\pi} 4\sin^2\frac{\omega}{2}d\omega ~\int\limits_0^\pi \sin\theta d\theta~\int\limits_0^{2\pi} d\phi \nonumber$.  
  We have also used the completeness relation of the $|\omega,\hat{w}\rangle$ basis.
     $z(C)$ is the eigenvalue of ${W}_C$ corresponding to the eigenstate $\prod_p |\omega_p,\hat{w}_p\rangle$. Since $W_C=\prod_{p_i} {\cal W}(p_i), ~z(C)=\prod_{p_i} z(p_i)$
 and $Tr z(C)= 2\cos\left(\omega(C)/2\right)$. Here, $\omega(C)$ is the gauge invariant angle characterizing the $SU(2)$ matrix $z(C)$ in its angle axis representation.   
Using the expression for the product of 2 SU(2) matrices \footnote{Product of 2 SU(2) matrices characterized by $(\omega_1,\hat{w}_1)$ and $(\omega_2,\hat{w}_2)$ gives an SU(2) matrix characterized by $(\omega,\hat{w})$ with  
%
\unexpanded{ 
\begin{align}
\cos\frac{\omega}{2} & = \cos\frac{\omega_1}{2}\cos\frac{\omega_2}{2}-(\hat{w}_1 \cdot \hat{w}_2) \sin\frac{\omega_1}{2}\sin\frac{\omega_2}{2}; \nonumber\\
~~~\hat{w}\sin\frac{\omega}{2}&=\hat{w}_1 \sin\frac{\omega_1}{2}\cos\frac{\omega_2}{2}+\hat{w}_2\sin\frac{\omega_2}{2}\cos\frac{\omega_1}{2} \nonumber\\ &\hspace{1.3cm}- [\hat{w}_1 \times \hat{w}_2] \sin\frac{\omega_1}{2}\sin\frac{\omega_2}{2}.
\end{align}}}
repeatedly, it is easy to show that 
%
%
$ \cos(\omega(C)/2) = \prod_i \cos(\omega_{p_i}/2)~ + $ terms which vanish on $\theta$ integration\footnote{The integrand under $\theta$ integration contains either $\sin2\theta$ or a $\cos\theta$, both vanish on $\theta$ integration from $0$ to $\pi$.} 
 Therefore, 
\begin{align}
\langle TrW_C\rangle
& = 2\left(\frac{I_2(2\alpha)}{I_1(2\alpha)}\right)^{n_c} 
= ~2e^{-n_c ~\ln \left(\frac{I_1(2\alpha)}{I_2(2\alpha)}\right)}
 \label{arealaw}
\end{align}
In (\ref{arealaw}), $n_c$ is the number of plaquettes in the loop C and $I_l(2\alpha)$ is the $l$-th order modified Bessel function of the first kind. We have used the relation \begin{align} I_l(2\alpha)= \frac{1}{\pi} \int\limits_0^\pi e^{2\alpha \cos\omega} \cos l\omega ~d\omega \label{bessel}\end{align} and the recurrence relation \cite{abrahamovich} \begin{align}I_{l-1}(2\alpha)-I_{l+1}(2\alpha) = \frac{2l}{2\alpha} I_l(2\alpha)\label{recur}\end{align} to arrive at (\ref{arealaw}).

The expectation value of the disorder operator in the variational ground state $|\psi_0\rangle$ is 
{\footnotesize
\begin{align}
&\frac{\langle \psi_0|\Sigma^\pm_{\theta=2\pi}(P)|\psi_0\rangle}{\langle\psi_0|\psi_0\rangle}= \frac{\prod\limits_{\bar{p}\neq P}~
	 {}_{\bar{p}}\langle\psi_0|\psi_0\rangle_{\bar{p}}~ {}_{P} \langle \psi_0|\Sigma^\pm_{2\pi}(P)|\psi_0 \rangle_P}{\prod\limits_{\bar{p}\neq P} {}_{\bar{p}}\langle\psi_0|\psi_0\rangle_{\bar{p}} ~ {}_{P}\langle\psi_0|\psi_0\rangle_{P}}\nonumber\\
&=\frac{\int d\mu(\omega_p,\hat{\omega}_p) e^{2\alpha [\cos(\frac{\omega_P+2\pi}{2})+\cos{\frac{\omega_P}{2}]}}}{\int d\mu(\omega_p,\hat{\omega}_p)e^{2\alpha\cos{\omega_P/2}}} ~~~=~~ \frac{\pi\alpha}{2I_1(2\alpha)}
\end{align}
}
Above, we have again used relations (\ref{bessel}) and (\ref{recur}) to get the last equality.

\section{Calculation of $\frac{\langle\psi_0|H_{spin}|\psi_0\rangle}{\langle \psi_0|\psi_0\rangle}$.}
\label{appc} 
%
The local effective SU(2) spin model Hamiltonian is  
\begin{align}
    \label{weakh2}
    H_{spin}=  \sum\limits_{p=1}^{\cal P} \left\{4g^2\vec{\cal E}^{~2}(p)+
    \frac{1}{g^2} \left[ 2- \left(Tr {\cal W}(p)\right)\right]\right\}+ \nonumber\\
    {g^2} \sum_{\langle p,p'\rangle}\left\{\vec{\cal E}_-(p)\cdot \vec{\cal E}_+(p')
    \right\} 
\end{align}
First, lets calculate $ \langle \psi_0|{\cal E}_-^a(p) {\cal E}_+^a(P)|\psi_0\rangle $. Here, $P$ is any plaquette.
\begin{align}
\label{eleexp1}
&\langle \psi_0|{\cal E}_-^a(p) {\cal E}_+^a(P)|\psi_0\rangle \nonumber\\
& = \Big\langle 0 \Big| \left( e^{S/2} {\cal E}_-^a(p) e^{-S/2}\right) e^S \left(e^{-S/2} {\cal E}_+^a(P) e^{S/2}\right) \Big|0\Big\rangle 
\nonumber\\ 
&= \frac{-1}{4} \big\langle \psi_0\big| \big[{\cal E}_-^a(p),S\big]\big[ {\cal E}_+^a(P), S \big] \big|\psi_0\big\rangle
\end{align}
In (\ref{eleexp1}), we have used the fact that ${\cal E}_\pm |0\rangle =0 $. Evaluating  $\langle \psi_0|{\cal E}_-^a(p) {\cal E}_+^a(P)|\psi_0\rangle$ in a different way, 
\begin{align}
\label{eleexp2}
& \langle \psi_0|{\cal E}_-^a(p) {\cal E}_+^a(P)|\psi_0\rangle \nonumber\\
& = \Big\langle 0 \Big|  e^{S/2} {\cal E}_-^a(p) e^{S/2} \left(e^{-S/2} {\cal E}_+^a(P) e^{S/2}\right) \Big|0\Big\rangle \nonumber
\end{align}
\begin{align}
&=  \frac{1}{2}  \big\langle \psi_0\big|\Big[{\cal E}_-^a(p),\big[ {\cal E}_+^a(P), S \big] \Big] \big|\psi_0\big\rangle \nonumber\\
& \hspace{1cm}+ \frac{1}{4} \big\langle \psi_0\big| \big[{\cal E}_+^a(P), S\big]\big[{\cal E}_-^a(p),S\big] \big|\psi_0\big\rangle
\end{align}
The equations (\ref{eleexp1}) and (\ref{eleexp2}) implies: 
%
%
\begin{equation}
\big\langle \psi_0\big|{\cal E}_-(p) \cdot {\cal E}_+(P)\big|\psi_0\big\rangle= \frac{1}{4} \big\langle \psi_0\big|\Big[{\cal E}_-^a(p),\big[ {\cal E}_+^a(P), S \big] \Big] \big|\psi_0\big\rangle
\label{pp}
\end{equation}
The expression in (\ref{pp}) vanishes when $P\neq p$. In particular,  
\begin{eqnarray}
\label{tevs} 
&~~~~~~\langle \psi_0|{\cal E}_-(p) \cdot {\cal E}_+(p')|\psi_0\rangle = 0, \\
& \langle \psi_0|{\cal E}_-(p) \cdot {\cal E}_-(p)|\psi_0\rangle = 
 \frac{3\alpha}{16}\langle \psi_0|Tr{\cal W}(p)|\psi_0\rangle.\nonumber 
\end{eqnarray}
%
%
%
Above $p,p'$ are nearest neighbors. Putting $n_c=1$ in equation (\ref{arealaw}), 
$\langle Tr{\cal W}(p)\rangle=\frac{2I_2(2\alpha)}{I_1(2\alpha)}.$
Using the above relations, the expectation value of $H_{spin}$ is
\begin{align}
\hspace{-0.4cm}\frac{\langle\psi_0|H_{spin}|\psi_0\rangle}{\langle \psi_0|\psi_0\rangle}= \hspace{-0.1cm} \sum\limits_p \left\{\hspace{-0.1cm} \left(\frac{3\alpha}{4}g^2-\frac{1}{g^2}\right) \frac{2I_2(2\alpha)}{I_1(2\alpha)}+\frac{2}{g^2}\right\}
\end{align}

The general non-local Hamiltonian $H$ differs from the above effective local spin Hamiltonian $H_{spin}$ by terms of the form $R_{ab}({W}) ~{\cal E}^a_-(p) {\cal E}^b_+(\bar{p})$, where $p$ and $\bar{p}$ are any 2 plaquettes on the lattice which are at least 2 lattice spacing away from each other. Above, ${W}$ is in general the product of many plaquette loop operators. 
The expectation value of the full Hamiltonian in the variational ground state $|\psi_0\rangle$ reduces to $\langle\psi_0|H_{spin}|\psi_0\rangle$ as the expectation value of the  non-local terms in $|\psi_0\rangle$ vanishes.

\end{document}